**Title**: Coarsening of biomimetic condensates in a self-stirring active fluid

**Authors:** Jeremy Laprade, Layne Frechette, Christopher Amey, Adrielle Cusi, Aparna Baskaran, W. Benjamin Rogers, Guillaume Duclos

Abstract

**Coarsening, the growth of larger structures at the expense of smaller ones, is a fundamental process in multiphase systems. The cell cytoplasm is a prototypical example of an out-of-equilibrium multiphase system, where molecular phase-separated condensates nucleate and grow in an active fluid composed of biopolymers and energy-consuming enzymes. Here, we uncover the mechanisms that govern the growth of condensates in a self-stirring active fluid by studying the coarsening of biomimetic condensates embedded within a 3D reconstituted cytoskeleton composed of microtubules and molecular motors. The remarkable agreement between experiments, an active hydrodynamic model, and a series of computer simulations reveals a comprehensive framework that explains the absence of self-similarity for active coarsening and the origin of the continuously varying coarsening exponents for both active and passive condensates. The coarsening dynamics is simply set by the statistics of droplets' binary collisions, which depend on their size-dependent motility, irrespective of its active or passive origins. These findings reveal a unifying control parameter for the coarsening and size distribution of active condensates, and expand our understanding of phase separation in out-from-equilibrium systems, with potential implications in materials science and biology.**

**Main**

Phase separation and coarsening are ubiquitous in multiphase complex fluids [1-6]. Phase separation generally occurs when a metastable homogeneous state demixes and evolves towards stable coexistence between two or more distinct phases. When a characteristic length scale is selected, the mixture forms a pattern. Otherwise, the system coarsens and the average size of the minority phase typically increases over time, reflecting the progressive ordering of the system as it reaches equilibrium. However, some multiphase mixtures coarsen without relaxing towards an equilibrium state. This is the case, for example, for passive fluids driven out-of-equilibrium by external forcing [7-9], or for active fluids whose components consume energy to drive chemical or chemo-mechanical reactions [10-12]. The cell cytoplasm is a prototypical example of an active mixture where the presence of energy-consuming enzymes breaks key assumptions from equilibrium thermodynamics [13]. While there are robust theoretical frameworks describing coarsening near equilibrium [14], analogous frameworks for active mixtures that are constantly consuming and dissipating energy are just emerging [15, 16].

Quenching a binary liquid into a metastable state can trigger the nucleation and growth of droplets of the minority phase, a process known as liquid-liquid phase separation (LLPS). Coarsening could result from i) Ostwald ripening, where differences in Laplace pressure favor the growth of large droplets at the expense of smaller ones [6, 14, 17], or ii) collision-driven coalescence—also known as Smoluchowski coagulation [18]—where motile droplets collide and merge to form larger droplets. Post nucleation, and without any external forcing, droplet coarsening follows power-law dynamics with a scaling exponent $\beta_R$ for the average radius R over time that depends on the growth mechanism [11, 14, 17, 19]. In the case of collision-driven coalescence in a passive Newtonian fluid and without any external forcing, the growth of the condensates is self-similar, i.e. time-independent when scaled by a time-dependent length, and the droplets' mean radius over time follows a power law with a $\beta_R = \frac{1}{3}$ exponent. The fundamental connection between the growth of self-similar patterns and the existence of a temporal power law for the average length scale of the system is the cornerstone of the classical theory of coarsening near equilibrium [14, 17, 20]. In contrast, under external shear, coarsening is suppressed and the binary mixture reaches a dynamical steady-state, both in 2D and 3D [7-9].

The mechanisms that control the coarsening dynamics of condensates *in vivo* remain to be fully elucidated, first because cells' interiors are not a simple Newtonian fluid, and second because cells are out-of-equilibrium. If the environment is Newtonian, motility scales as the inverse of the droplet's radius. When the environment is complex, additional scale dependence on the droplets' motility has to be taken into account. For example, in the cell nucleus, the presence of chromatin, a dense visco-elastic network composed of DNA and proteins, restricts the motion of condensates to subdiffusive dynamics, which effectively reduces their coarsening exponents [21]. Finally, the current framework for describing coarsening in cells often assumes that the demixed state relaxes to equilibrium. However, this is not necessarily true in living matter, given that cells actively consume energy to maintain themselves out-of-equilibrium.

Molecular condensates often nucleate and grow in the cytoplasm, a prototypical example of an active self-stirring fluid. The cytoplasm is driven out-of-equilibrium in parts by molecular motors that push and pull on the cytoskeleton. As a result, the mixture can spontaneously flow and advect condensates. Sometimes, condensates can even self-propel because they are composed of, or covered with, energy-consuming enzymes [12, 22, 23]. Active transport might impact condensate formation and dynamics in many biological settings, given that active flows are ubiquitous across the tree of life. In animals, active flows trigger symmetry breaking in C. Elegans embryos [24], control mitotic spindle positioning in mouse oocytes [25], and ensure mitochondrial homeostasis during interphase [23]. In plants, active flows enable intracellular cargo transport over long distances [26, 27] and promote subcellular compartmentalization [28]. Theory and experiments with reconstituted biomimetic materials have recently brought new insights into how active hydrodynamics might impact phase separation [29, 30], coarsening [31-33], wetting [34], and interfacial dynamics [35-37]. However, a comprehensive framework that describes how condensates embedded in an active fluid grow and coarsen over time is still lacking. The goal of this article is to fill that gap.

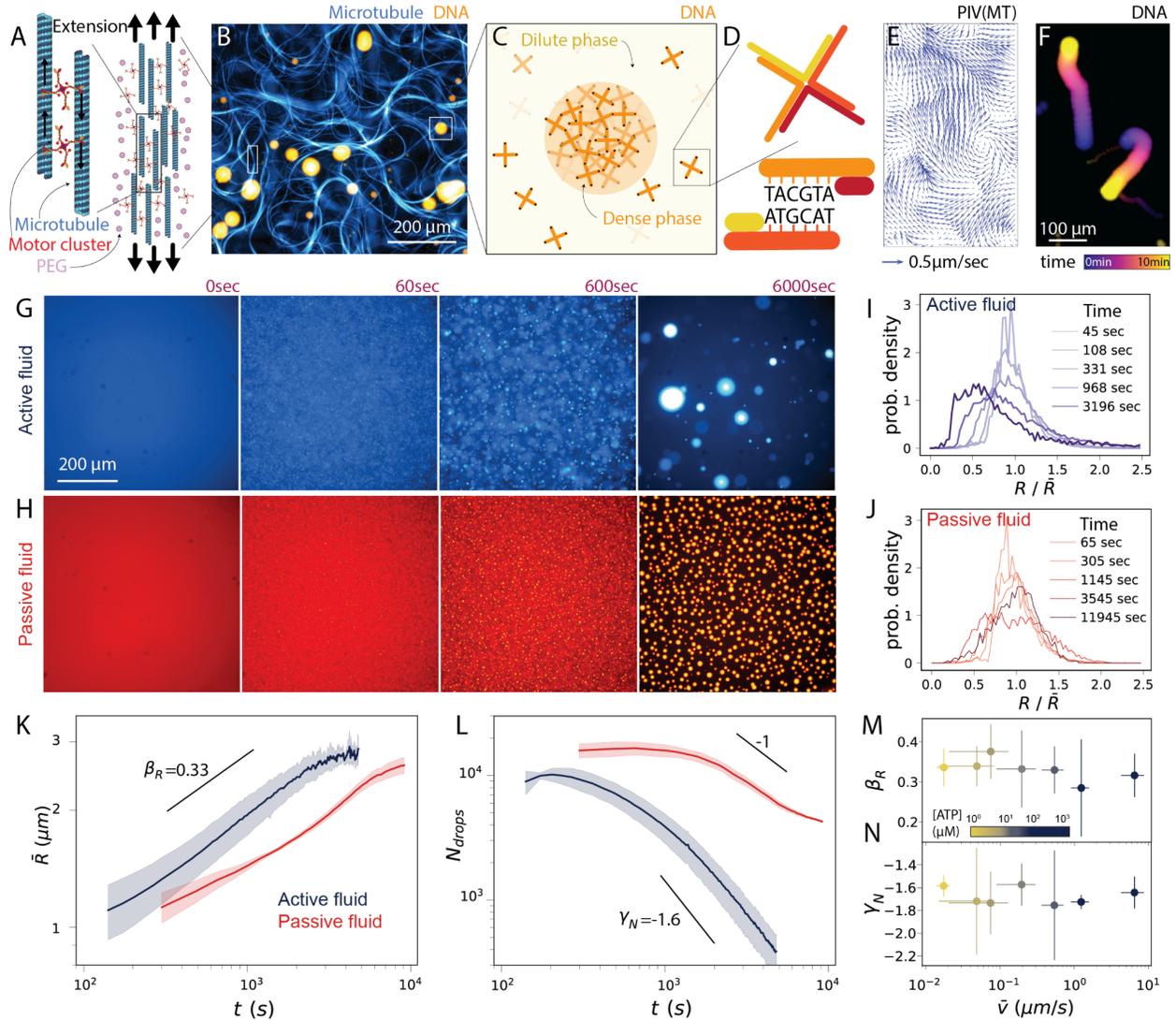

**Fig. 1: Coarsening of DNA droplets in an active self-stirring fluid composed of microtubules and clusters of molecular motors.** A) Schematic of the extensile kinesin-microtubule bundles. B) Fluorescent image where microtubules are labeled in cyan and DNA in orange. C) Schematic of a DNA-rich phase-separated droplet in a DNA-poor background. D) schematic of a DNA nanostar with its palindromic sticky ends. E) Instantaneous velocity field of the microtubules measured by Particle Image Velocimetry. F) Time-colored images of the DNA droplets for the same field of view as shown in E. The droplets are advected by the active flows generated by the microtubule bundles. G-H) Fluorescent time series of DNA droplets coarsening in G) a self-stirring active fluid or H) a passive fluid. I-J) Probability distribution function of the droplet's radius R renormalized by its mean R̄ for droplets in I) an

**The coarsening of DNA droplets in a 3D active self-stirring fluid follows a power law that is independent of the composition of the active fluid.**

Here, we study the coarsening dynamics of phase-separating DNA nanostars [38] embedded within a 3D self-stirring isotropic active fluid confined between two parallel plates and composed of extensile microtubule bundles, a depletant (20kDa PEG), and clusters of molecular motors [39] (**Fig. 1A-D, Movie S1**). The motor clusters hydrolyze ATP, driving interfilament sliding and microtubule bundle extension (**Fig. 1A**). Quenching the composite fluid to room temperature leads to the nucleation and growth of DNA-rich droplets in a DNA-poor active fluid (**Fig. 1G**). At room temperature, the DNA-rich droplets remain in a liquid phase, continuously exchanging material with the surrounding DNA-poor phase [40, 41]. The depletant responsible for microtubule bundling diffuses within the DNA-rich droplets [42], and therefore does not contribute to the coarsening process. The experimental sample chamber is about 500 times longer (~3 mm) than the active length scale (~60 μm), which is set by the confinement along the Z-axis. The extensile motion of the microtubule bundles generates chaotic flows with a steady speed and a steady active length scale that continuously advect the DNA condensates (**Fig. 1E-F, Movie S2, Fig. S1**). After the volume fraction of the DNA-rich phase reaches a steady state (**Fig. S2**), growth of the droplets results primarily from Smoluchowski coarsening and not from Ostwald ripening. We measured that the coarsening rate from Ostwald ripening of immobile droplets is significantly slower than the growth resulting from colliding droplets (**Fig. S2**). The droplets' size grows as two small droplets collide and fuse into a single large droplet (**Fig. S3**). The droplets are not constrained by the percolating microtubule network (**Fig. S4**). In what follows, we compare the coarsening of active droplets - droplets advected by an active fluid - and the coarsening of passive droplets - droplets diffusing in a passive fluid that does not contain any microtubules (**Fig. 1H**).

We segmented the DNA-rich droplets and measured their number, size distribution, and the first and second moments of this size distribution over time (See Materials and Methods, **Fig. S5**). Measuring the probability distribution function of the active droplets' radii R normalized by their mean R̄ revealed that a large number of small droplets coexist with a few exceptionally large ones (**Fig. 1I**). This asymmetry in the active droplet's relative size distribution increased over time (**Fig 1I, Fig. S6**), while the relative size distribution of the passive droplets did not change (**Fig. 1J, Fig. S6**). Both active and passive droplet size distributions were well-fitted by a log-normal distribution. We measured $\beta_R$, the scaling exponent for the temporal evolution of the mean radius <R(t)>, and $\gamma_N$, the scaling exponent for the temporal evolution of the total number of droplets N(t) (See Materials and methods for the fitting procedure). The $\beta_R$ exponent was the same for active and passive coarsening when DNA-rich droplets diffuse in a passive fluid with no microtubules ($\beta_{R\ Active}$=0.33+/-0.03, $\beta_{R\ Passive}$=0.32+/-0.04 , **Fig. 1K**). The $\gamma_N$ exponent was smaller for active than for passive coarsening ($\gamma_{N\ Active}$=-1.65 +/- 0.08, $\gamma_{N\ Passive}$=-0.89 +/- 0.04,

**Fig. 1L**). The two exponents for active coarsening were robust: we consistently measured the same exponents irrespective of the concentrations of ATP or confinement along the Z-axis, which respectively control the mean speed and the active lengthscale (**Fig. 1M-N, Fig S7**).

Comparing the coarsening of DNA droplets in an active and a passive fluid reveals some striking differences and some less intuitive similarities. First, the exponents and size distributions of passive DNA droplets are consistent with Smoluchowski coarsening for Brownian diffusion ($\beta_{R\ Smolu}$=⅓, $\gamma_{N\ Smolu}$=-1) [14, 43]. Second, coarsening in an active fluid is greatly accelerated compared to the passive fluid (**Fig. 1G-H**). Third, contrary to passive coarsening, active droplet coarsening is not self-similar: the relative size distribution of the active droplets evolves over time, while the one for passive droplets is time-independent. Still, the mean-radius dynamics follow the same scaling law for active and passive droplets, which challenges the fundamental connection from the classical coarsening theory between self-similarity and the existence of a temporal power law for the mean radius. The total number of droplets, however, exhibits different exponents for active and passive coarsening. Finally, active coarsening follows exponents that do not depend on the rates of the underlying transport phenomena. The goal of this paper is to uncover the origin of the loss of self-similarity and the emergence of the two power-law exponents $\beta_R$ and $\gamma_N$ for active coarsening.

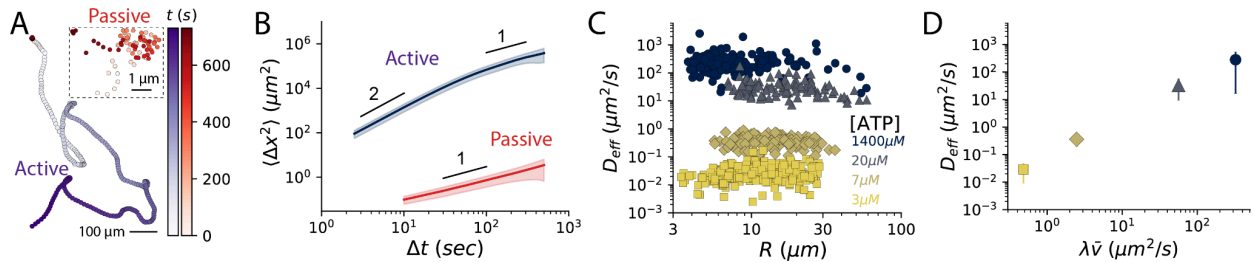

**Fig. 2: DNA droplet transport in a self-stirring active fluid.** A) Time-colored trajectories of DNA droplets in an active fluid (purple) or a passive fluid (red). Note the difference in scale bars. B) Mean Square Displacement of active and passive droplets. C) Effective diffusivity for DNA droplets of increasing size embedded in active fluids with varying concentrations of ATP. D) Mean effective diffusivity as a function of the product of the mean speed of the active fluid with the mean velocity-velocity correlation length λ. Data points are color-coded depending on the ATP concentration. Data points and error bars show the mean and standard deviation of the mean for at least N=102 droplets.

**DNA droplet transport in a 3D self-stirring active fluid.**
Given that coarsening results from the collision of motile droplets, we first characterized their motion when immersed in an active fluid. We tracked DNA droplets of various sizes immersed in a steadily self-stirring active fluid or a passive fluid (**Fig. 2A**). The active droplets exhibited ballistic motion at short-time scales and diffusive-like motion at longer time scales (**Fig. 2B**). The length scale associated with the ballistic to diffusive crossover (~ 60 μm) is significantly shorter than the linear dimensions of the sample chamber (~3 cm). Droplets in a passive fluid were diffusive and explored much smaller areas relative to the droplets in an active fluid (**Fig. 2B**). Measuring the effective diffusivity of DNA droplets of various sizes revealed a fundamental difference between active and passive coarsening: the effective diffusivity of the droplets in the active fluid was independent of the droplets' sizes (**Fig. 2C**). In other words, the Stokes–Einstein–Sutherland (SES) relation, which imposes that, at low Reynolds number, diffusivity and radius are inversely proportional for passive droplets (**Fig. S8**), was not valid for the effective diffusion of droplets advected in an active fluid. Finally, the properties of the active flows set the effective diffusivity, which was equal to the product of the velocity autocorrelation length and the average speed of the active flows (**Fig. 2D**). Changing the molecular composition of the active fluid or vertical confinement did not change the exponents of the droplets' mean squared displacement. Instead, it impacted the speed and the characteristic vortex size of the active flows, and therefore the effective diffusivity of the DNA droplets (**Fig. 2D**).

At this point, we want to highlight two fundamental observations. First, measuring similar coarsening exponents $\beta_R$ for active and passive droplets (**Fig. 1K**) is not intuitive given that i) active droplets do not follow the SES relation, while ii) the SES scaling exponent plays a key role in setting the passive scaling exponents [43]. Second, the long-term diffusive-like motion of the active droplets results from the advection by the chaotic active fluid whose flows have a finite spatial and temporal correlation (**Fig. S1**). The active diffusivity has nothing to do with Brownian diffusion, and a size-dependent effective kinetic temperature would not be thermodynamically meaningful, given that the droplets' motion does not result from thermal effects but from size-independent active advection.

**Coarsening dynamics in computer simulations of active and passive transport.**
We performed simulations of droplet coarsening in a self-stirring active fluid and in a passive fluid to elucidate the mechanisms that set the value of the coarsening exponents. We solved the active nematohydrodynamic equations in 3D using a multigrid solver (see Materials and Methods) and measured the coarsening dynamics of droplets advected by the active flows (**Movie S3, Fig. 3B, Fig. S9**). The results of the hydrodynamic simulations agreed remarkably well with the experimental coarsening dynamics for DNA droplets in an active fluid: the size distribution of active droplets was not self-similar (**Fig. 3D**), active flows accelerated coarsening, and the coarsening exponents $\beta_R$ and $\gamma_N$ had the values consistent with the experimental results ($\beta_{R\ Active\ Hydro}$= 0.30 +/- 0.07 , $\gamma_{N\ Active\ Hydro}$=-1.61 +/- 0.07, **Fig. 3E-F, H**).

To investigate the origin of these exponents, we performed additional computer simulations in which we programmed various types of droplet motility. In particular, we measured the coarsening dynamics of i) self-propelled droplets, ii) diffusive droplets following a modular Stokes-Einstein-Sutherland relation (D~$R^{-m}$, with an SES exponent m varying between -1 and 6), and iii) droplets undergoing anomalous diffusion (MSD~D*$t^\alpha$ where α is the anomalous diffusion exponent, α<1 for subdiffusive droplets, α=1 for diffusive droplets, and α>1 for superdiffusive droplets).

We first investigated the coarsening of self-propelled droplets. Active Ornstein-Uhlenbeck Particles (AOUPs), a common model for active particles, execute persistent random walks (**Fig.**

**S10**). AOUPs exhibit trajectories and mean-square displacements that are comparable to those of droplets embedded in an active fluid, but lack the spatial correlations induced by hydrodynamics. We measured similar coarsening exponents for AOUPs as for droplets advected by an active fluid (**Fig. 3H, Fig. S10E-F**). Measuring the coarsening dynamics of other types of self-propelled particles— Active Brownian Particles (ABPs) and Run-and-Tumble particles (RTPs)—revealed that the values of the coarsening exponents are robust and do not depend on the details of the self-propulsion mechanism or on the persistence time of the particles (**Figs. S10, S11**). These results also demonstrate that the active coarsening exponents are not sensitive to spatial correlations among droplet trajectories.

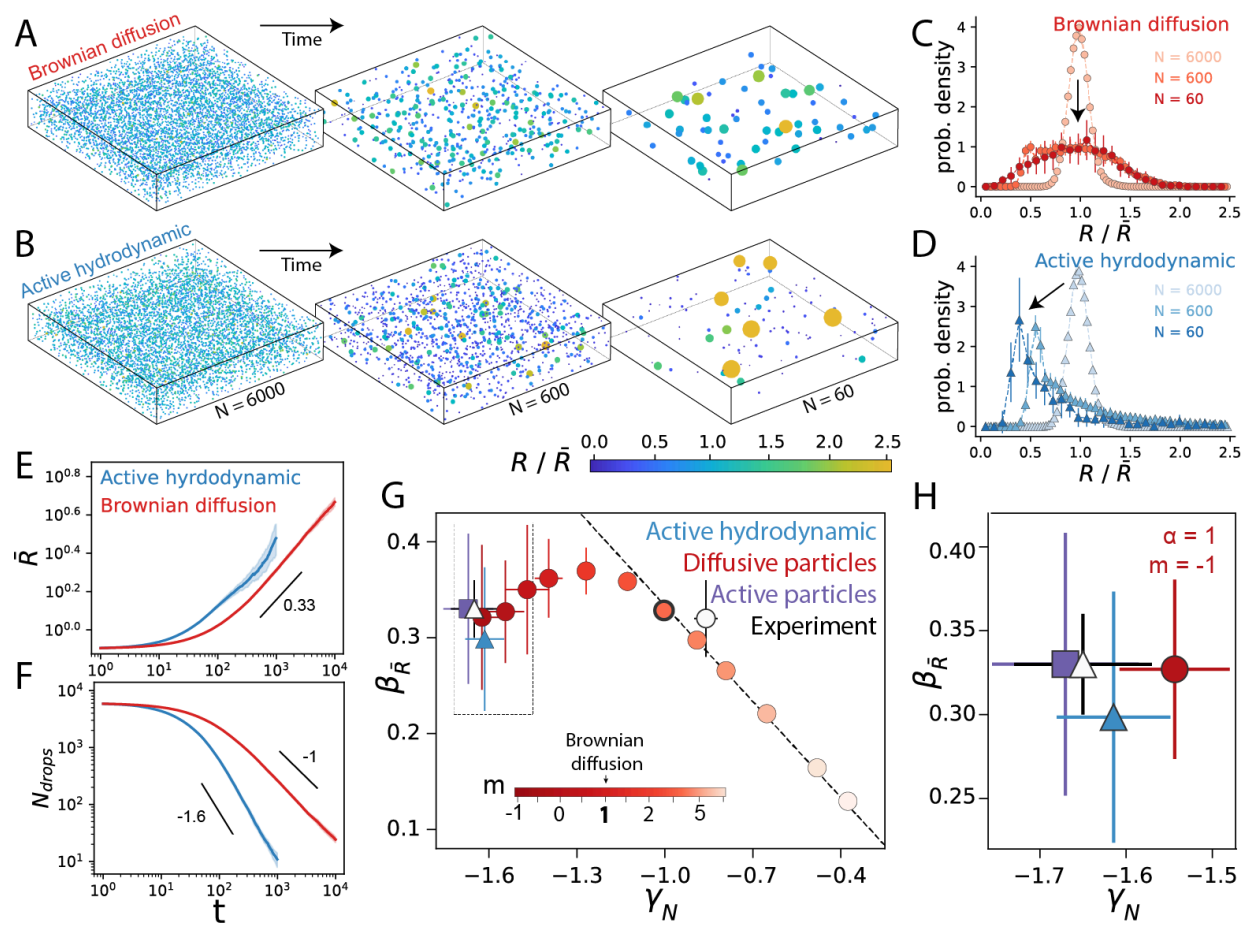

**Fig. 3: Coarsening dynamics in computer simulations of active and passive transport.**
A-B) Time series of coarsening simulations for A) Brownian droplets in a passive fluid and B) droplets advected by a 3D active fluid. Droplets are color-coded according to their radius R relative to R̄, the mean radius at that simulation step. Simulation frames in each column have the same number of droplets, N, but different time stamps. C-D) Probability distribution function of the radius of droplets embedded in C) a passive or D) an active fluid. E) Temporal evolution of the mean radius of the droplets. F) Temporal evolution of the total number of droplets N. G) $\beta_{\bar{R}}$ and $\gamma_N$ coarsening exponents for diffusive droplets with a modular Stokes-Einstein-Sutherland's exponent m (D~R$^{-m}$, with m=-1:7). Brownian particles (m=1) have a bold marker. For m<m*=1, $\beta_{\bar{R}}$ deviates from - $\gamma_N$/3 (dashed line). The coarsening exponents for We next investigated the coarsening of diffusive droplets with a modular Stokes-Einstein-Sutherland relation ($D \sim R^{-m}$, **Movie S4**). Decreasing the SES exponent m revealed a non-linear relationship between the two coarsening exponents $\beta_R$ and $\gamma_N$ (**Fig. 3G**). $\gamma_N$ increased monotonically, while $\beta_R$ was not monotonic with $m$ (**Fig. S12**). Furthermore, $\beta_R$ deviated from the expected $\beta_R$=-$\gamma_N$/3 linear relationship for the coagulation of Brownian droplets as soon as $m$ was smaller than m*=1, its expected value from Stokes-Einstein-Sutherland at thermodynamic equilibrium. m* was also the threshold value of the exponent $m$ below which the droplets' size distributions lost their self-similarity (**Fig. S13**). When m is negative, larger droplets diffuse faster than smaller droplets. Diffusive droplets with an inverted Stokes-Einstein-Sutherland relation ($D_m \sim R$, m=-1) had similar coarsening exponents as the droplets in an active fluid (**Fig. 3H**).

We then performed coarsening simulations for anomalously diffusing droplets that undergo a continuous-time random walk with a modular Stokes-Einstein-Sutherland relation (MSD$\sim D^* t^\alpha$ with $0 < \alpha < 2$, see materials and methods). The coarsening dynamics exhibited continuously varying exponents and revealed a non-linear relationship between $\beta_R$ and $\gamma_N$ (**Fig. S14**). The transition from self-similar to anomalous "active-like" coarsening and the loss of self-similarity emerged for different values of m*, depending on the anomalous diffusion exponent $\alpha$. Subdiffusive (resp. superdiffusive) droplets lost their self-similarity and $\beta_R$ deviates from -$\gamma_N$/3 for smaller (resp. larger) values of m (**Fig. S15**).

Taken together, these simulations revealed that the coarsening exponents $\beta_R$, $\gamma_N$, and their relationship are fundamentally linked to the droplets' motility, irrespective of its active or passive origin. The mechanistic connection between motility and coarsening exponents will become clear once we look at the binary collision dynamics. Interestingly, different combinations of the anomalous diffusion exponent $\alpha$ and the Stokes-Einstein-Sutherland exponent m can lead to the same coarsening power laws. For example, both superdiffusive condensates and the preferential attachment effect in inverted Stokes-Einstein diffusion promote the non-self-similar coarsening seen in active condensates. Finally, we see that the loss of self-similarity is accompanied by a deviation from the linear relationship $\beta_R$=-$\gamma_N$/3. We then looked beyond single-particle motility and characterized binary collision dynamics for active and passive condensates to further unify these results and reveal why different types of motility exhibit similar coarsening dynamics.

**Collision kernels reveal a unifying control parameter for active and passive coarsening.**
We quantified the binary collision rates between motile droplets of various sizes to investigate the origin of the non-linear relationship between the two coarsening exponents $\beta_R$ and $\gamma_N$. Specifically, we performed additional simulations of self-propelled droplets where we registered droplet collisions but disabled droplet coalescence (see Materials and Methods). These simulations enabled us to measure a collision kernel $\Gamma_{i,j}$—sometimes called the coagulation kernel, or the condensation kernel—a probabilistic map of the binary collision rates for droplets of sizes $R_i$ and $R_j$. We first confirmed that we recover the collision kernel $\Gamma_{i,j}=4*\pi*(1/R_i+1/R_j)*(R_i+R_j)$ predicted by Smoluchowski for Brownian diffusion ($\alpha=1$, $m=1$, **Fig. 4A**). We found that the most probable collisions occurred for droplets of dissimilar radius: large droplets that diffuse slowly are easy to find because of their large capture cross-section, while small droplets diffuse quickly and therefore explore space very efficiently.

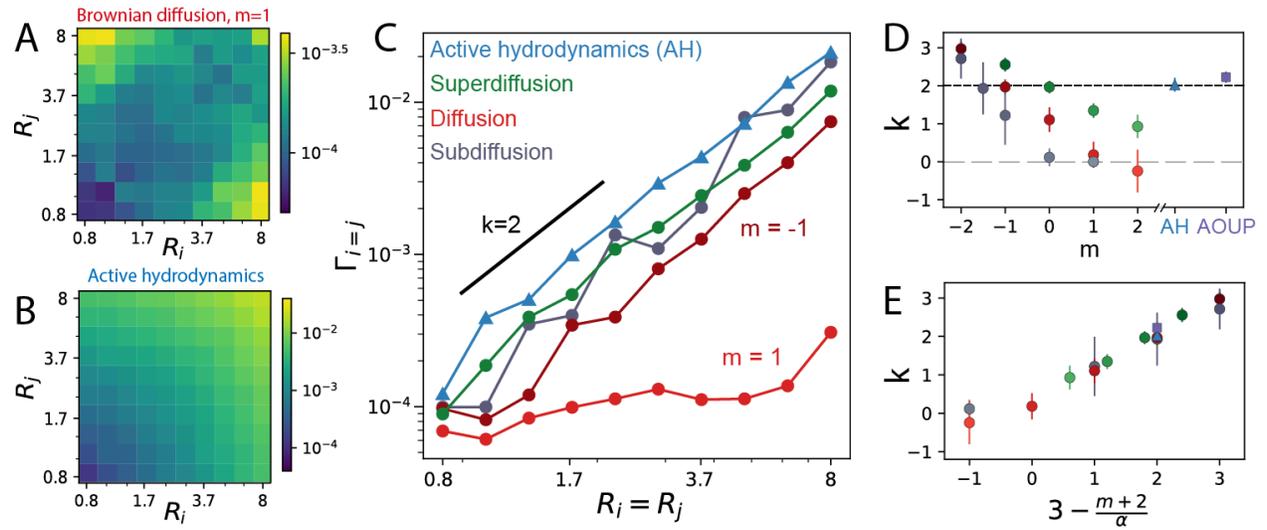

**Fig. 4: Collision kernels reveal a unifying control parameter for active and passive coarsening.** A) Collision kernel for Brownian particles ($\alpha=1$ and $m=1$). $\Gamma_{i,j}$ is the probability that droplets of size $R_i$ and $R_j$ collide. B) Collision kernel for particles embedded in a self-stirring active fluid. C) Collision probability as a function of droplet radius for pairs of same-sized droplets. k is the scaling exponent of the collision kernel $\Gamma_{i=j}\sim R_i^k$. D) Scaling exponent k of the collision kernel $\Gamma_{i=j}$ for motile droplets with different motility: diffusive ($\alpha=1$, red curves), subdiffusive ($\alpha=0.5$, grey curve), superdiffusive ($\alpha=1.67$, green curve), as well as for droplets embedded in a self-stirring active fluid (blue triangle) and for self-propelled droplets (Active Ornstein-Uhlenbeck Particles, AOUPs, purple square). Data points and error bars show the mean and standard deviation of the mean for N=12 independent replicates. E) Scaling exponent k of the collision kernel as a function of 3-(m+2)/$\alpha$, where m is the scaling of the Stokes-Einstein-Sutherland relation and $\alpha$ is the anomalous diffusion exponent. All the data points collapse on the same line. The color code is identical to the data shown in panels C and D.

The collision kernel for droplets immersed in an active fluid differed drastically from the kernel obtained for passive Brownian droplets. The most probable collisions occurred between the largest active droplets (**Fig. 4B**): they have the largest cross-section and explore space as effectively as smaller droplets given that the active advection is independent of droplet size (**Fig. 2C**). Measuring the collision kernel for Active Ornstein-Uhlenbeck Particles, diffusive droplets with an inverted SES relation ($\alpha=1$, $m=-1$), and superdiffusive droplets ($\alpha=1.7$, $m=0$) reveal the existence of a unifying control parameter that explains why these distinct motility mechanisms exhibit the same coarsening exponents (**Fig. S16**). All of these collision kernels $\Gamma_{i,j}$ displayed the same scaling exponents $k=2$ with droplet radii $R_i$ and $R_j$: $\Gamma_{i,j} \sim (R_i+R_j)^k$ (**Fig. 4C**).

Systematically measuring the kernel exponent k for diffusive and anomalously diffusive droplets revealed a fundamental relationship between the collision kernels' exponent k, the anomalous diffusive exponent $\alpha$, and the Stokes-Einstein-Sutherland's exponent m (**Fig. 4D-E**): $k=3-(m+2)/\alpha$.
Any value of k greater than 0 favors collisions between two large droplets at the expense of droplets with more dissimilar sizes. This phenomenological relationship explains why the critical value $m^*$ of the SES exponent below which the coarsening of anomalously diffusing droplets loses self-similarity depends on the MSD exponents $\alpha$ (**Fig. S13**). The coarsening dynamics of superdiffusive and subdiffusive droplets stopped exhibiting self-similarity as soon as their motility favored the collision between the largest droplets ($k>0$). Indeed, the preferential coalescence between the largest droplets leads to the exacerbated coexistence between a few exceptionally large droplets that constantly coalesce and many small droplets that rarely coalesce. The asymmetry in the relative size distribution increases over time, hence the absence of self-similarity.

**Probabilistic calculation of the coarsening exponent $\beta_R$.**
To further confirm the relevance of the scaling exponent of the collision kernel on the coarsening dynamics, we used an iterative probabilistic method to calculate how the mean droplets' radius evolves for a prescribed collision kernel and a prescribed collision rate exponent $\gamma_N$. Starting from a set of N droplets with a given size distribution, we merged at each time step a number of droplets set by N and $\gamma_N$. The pairs of merging droplets were chosen from a 2D probability distribution set by the collision kernel and the droplet size distribution. We recalculated the new size distribution at each time step and performed the calculation until all the droplets merged into a single droplet.

Using this approach, we first evaluated the coarsening dynamics in simple limit scenarios. We note that enforcing that the smallest droplets collide at every timestep led to a $\beta_R$ exponent similar to the one for diffusive droplets (**Fig. 5A**, black solid line). Interestingly, enforcing instead that the largest droplets collide at every timestep led to, at first, a decrease in the mean droplets' radius over time (**Fig. 5A**, black dashed line). We also measured the expected coarsening exponent $\beta_R=\frac{1}{3}$ for the Brownian coagulation kernel ($\Gamma_{i,j}=4*\pi*(1/R_i+1/R_j)*(R_i+R_j)$, $k=0$) and recovered the linear relationship $\beta_R=\gamma_N/3$ (**Fig. 5B** for $k=0$), a well as the self-similar probability distribution function for the droplet's radiuses (**Fig. S17**), validating our approach.

Motivated by the scaling of the collision kernels measured for active and anomalously diffusing particles, we calculated the coarsening dynamics for droplets whose collision dynamics are determined by the following kernel: $\Gamma_{i,j}=(R_i+R_j)^k$. Here, we found that the relationship between $\beta_R$ and $\gamma_N$ became non-linear for positive collision kernel exponents k (**Fig. 5B, Fig. S18**). For a given $\gamma_N$, increasing k led to a decrease of $\beta_R$. In other words, the non-linearity between $\beta_R$ and $\gamma_N$ grew as the collision kernel exponent k increased. This calculation recovered the exponents measured for active and passive coarsening simulations (**Fig. 5B inset**). Furthermore, increasing the collision kernel exponent k led to a loss of self-similarity for the droplet's radius probability distribution function (**Fig. 5C-D**), with a log-normal distribution function that became more asymmetric as k increases (**Fig. 5D**).

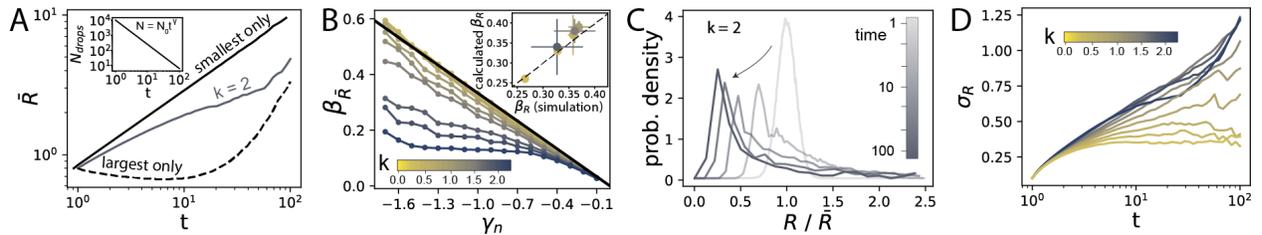

**Fig. 5: Probabilistic calculation of coarsening exponents.** A) Temporal evolution of the mean radius of the droplets for a prescribed $\gamma_N$= -1.6 exponent and a given coalescence protocol: only the smallest droplets merge at each step (black solid line), only the largest droplets merge (black dash line), or pairs of droplets chosen from the probabilistic collision kernel merge (gray solid line for a collision kernel exponent k=2). The inset shows the prescribed temporal evolution of the total number of droplets. **Fig. S16** shows the temporal evolution of the mean radius for other values of k. B) Plot of the two coarsening exponents $\beta_R$ as a function of $\gamma_N$ for increasing collision kernel exponent k. For k=0, $\beta_R=\gamma_N/3$ (continuous back line). The inset shows the value of $\beta_R$ obtained with the probabilistic calculation versus the value of $\beta_R$ measured from the coarsening simulations in Fig. 3H. The dashed line shows the identity function. C) Temporal evolution of the probability distribution of the droplets' radius for a collision kernel exponent k=2. D) Temporal evolution of the standard deviation \sigma_R of the radius probability distribution function for increasing values of k.

Taken together, these results demonstrate that the positive scaling exponent of the collision kernel is the fundamental control parameter that dictates the non-linear relationship between the two coarsening exponents $\beta_R$ and $\gamma_N$. Different types of droplet motility can exhibit the same collision kernels, which explains why droplets embedded in an active self-stirring fluid, self-propelled droplets (AOUPs, ABPs, RTPs), and anomalously diffusing passive droplets with various Stokes–Einstein–Sutherland scalings exhibit the same coarsening exponents. This calculation shows that the coalescence of a pair of large droplets does not grow the mean radius as effectively as the coalescence of a pair of smaller droplets because R scales as $V^{1/3}$ and volume is conserved during coalescence (**Fig. S2**). Second, this calculation demonstrates that while active coarsening is the most effective way to quickly grow the largest droplet, passive coarsening with Brownian diffusion is more effective at increasing the average size of the condensate population. Finally, while subdiffusive droplets have a narrower size distribution compared to Brownian droplets, active condensates exhibit broader size distributions which could have important physiological implications since large size variations, seen in nuclear condensates like nucleoli [44] and Cajal bodies [45], are linked to various diseases, including cancer [46] and premature aging [47].

**Conclusion and Discussion**

In conclusion, our results reveal the origin of the continuously varying coarsening exponents for active and anomalously diffusing passive droplets with varying size-dependent motility. We conclude by noting a few limitations of our study that may guide future experimental work. First, a primary open question concerns how the observed scaling behavior persists as the system size is varied. In our experiments, the active system is approximately 500 times larger than the active length scale, which we take as a proxy for the crossover length where droplet motion shifts from ballistic to diffusive. Exploring this scaling in both larger and smaller systems could help clarify the extent to which our results reflect the thermodynamic limit or finite-size effects. Investigating coarsening in smaller systems may also offer insights more directly applicable to confined, cell-like environments. Second, It is important to note that the droplet sizes examined here remained below the active length scale, suggesting the possibility of a crossover to a different coarsening regime as droplets size approach this scale. Droplets larger than this lengthscale might experience a different mode of transport, or significant shearing. To note, previous studies for passive droplets have demonstrated that coarsening can be suppressed under external shear flow [7-9]. Future studies should investigate the coarsening dynamics of larger condensates in active fluids. Finally, in this regime, the role of Ostwald ripening may warrant reevaluation [14, 17], particularly given the broad distribution of droplet sizes that emerges during coarsening in active systems.

We now turn to the biological relevance of our findings, in particular regarding nuclear condensates. Previous studies have shown that nuclear condensate coarsening is primarily driven by collision-induced coalescence rather than Ostwald ripening [21, 48]. While self-stirring dynamics—characteristic of biomimetic active fluids—are typically more relevant in large cells undergoing cytoplasmic streaming [25-28], there is some evidence that such flows can also occur within the nucleus [49]. Of note, the results presented here are valid when the system size exceeds the characteristic length scale over which droplets exhibit diffusive motion, a condition that may not be met in small cells. Nevertheless, our theoretical framework could apply to certain in vivo scenarios, even in the absence of active flows—for instance, during the coarsening of nuclear condensates. There, chromatin mechanics leads to subdiffusive condensate motion with an effective diffusivity that deviates from the classical Stokes–Einstein–Sutherland relation [21]. Under such conditions, we estimate the coarsening kernel exponent k=0, yielding a linear relationship $\beta_R = -\gamma_N/3$, in agreement with previous observations of nuclear condensate coarsening [21]. Second, coalescence-driven coarsening in pathological situations where condensates continuously nucleate has been reported to be consistent with preferential attachment [48], implying a collision kernel with a positive exponent k>0. It would be interesting to compare our predictions with experimental measurements of the coarsening exponents, the Mean Square Displacement, and the collision kernels of slowly nucleating cytoplasmic condensates such as Huntingtin polyQ exon 1 [48], a protein implicated in Huntington's disease [50], a progressive neurodegenerative disorder. Observing power laws consistent with our predictions would suggest that our findings are broadly applicable, not only to self-propelled or actively transported condensates but also to other biological condensates that exhibit preferential attachment.

The comprehensive framework presented in this article allows us to disentangle the respective contributions of active flows and passive mechanical processes to the coarsening dynamics of molecular condensates. This decoupling is crucial given the ubiquity of LLPS, and both active and passive mechanics across diverse biological systems, from active transport in oocytes and cytoplasmic streaming in plant cells to enzymatic activity in the nucleus. Finally, this framework is generalizable to volume-conserving coarsening in various flow environments and may offer insights into other natural processes, such as platelet aggregation during blood clot formation under laminar flows [51], or the agglomeration of nanoplastics in turbulent oceanic currents [52].

**Acknowledgment**

This work was supported by a NSF CAREER award DMR-2047119 (J.L, A.C., and G.D) and the NSF-funded Brandeis Bioinspired MRSEC DMR-2011846 (L. F., C.A., and A.B.) We thank Dr. Shibani Dalal, director of the Brandeis Biomaterial Facility, for help with protein purification. We also acknowledge the use of the optical, microfluidics, and biomaterial facilities supported by NSF MRSEC DMR-2011846. W.B.R. acknowledges support from the Human Frontier Science Program (RGP0029).

**Materials and methods content:**
A. Protein biochemistry
B. Microscopy and image analysis
C. Details about the computer simulations
D. Supplementary Figures
E. Supplementary movie description

**Materials and methods**

**Title**: Coarsening of biomimetic condensates in a self-stirring active fluid

### A. Protein biochemistry

*Microtubules*

We purified tubulin from calf brains. Upon receiving the brains (generally three to four brains, total mass ~ 1kg), we immediately cleared fat and blood clots from the materials. Brains were not frozen or stored in any way before the purification process. We liquefied clean brain materials in 50 mM 2-(N-morpholino)ethanesulfonic acid (MES), 1mM $CaCl_2$ buffer (roughly 1 L/kg of brain). We then centrifuged the material at 10,000 rpm for 45 minutes to remove large solids from the sample. We performed a secondary spin (10,000 rpm for 20 min) for further clarification of the supernatant. After recovering the supernatant, we performed a two-step polymerization and depolymerization protocol to improve tubulin purity. To polymerize microtubules, we incubated the tubulin at 37°C in a 1M piperazine-N,N′-bis(2-ethanesulfonic acid) (PIPES) buffer with 100 µM adenosine triphosphate (ATP, Sigma Aldrich A1852), 200 mM guanosine-5'-triphosphate (GTP, Sigma Aldrich Prod. No. G8877), and ⅓ volume glycerol for one hour. After polymerization, we centrifuged the polymerized tubulin at 37°C, 44,000rpm for 30 minutes. We recover the pellet containing microtubules and discard the supernatant. We then depolymerize the microtubule pellets in the MES buffer at 4°C for 30 minutes. For further homogenization, we used a dounce homogenizer for 30 minutes. We then centrifuge the depolymerized tubulin at 30,000 rpm for 30 minutes at 4°C. The final depolymerization step is done in an 80mM PIPES buffer. We measured the tubulin concentration using a nanodrop spectrophotometer (generally around 20mg/mL), aliquoted, flash froze, and stored the tubulin at -80°C.

We assembled microtubules from recycled tubulin. We mixed 80 µM tubulin (3% labeled with Alexa 647), 0.6 mM GMPCPP (Fisher Scientific Cat. No. NC1303776), 1 mM dithiothreitol (DTT, Fisher Scientific Cat. No. AC165680050), and 80 mM PIPES buffer. We incubated the mixture at 37°C for 30 minutes. We then let the microtubule mixture rest at room temperature (25°C) for five hours. We flash-froze and stored the polymerized microtubules at -80°C for use in future experiments.

*Kinesin Motor Clusters*

K401- BIO-6xHIS (processive motor, dimeric MW-110 kDa) corresponds to the 401 amino acids derived from N-terminal domain of Drosophila melanogaster kinesin-1, and labeled with 6-histidine and a biotin-tag [84, 85]. The motor proteins were transformed and expressed in Rosetta (DE3) pLysS cells and purified following protocols described elsewhere [84]. The purified proteins were flash-frozen in liquid nitrogen with 36% sucrose and stored at -80 ∘C. We used tetrameric streptavidin (ThermoFisher, 21122, MW: 52.8 kDa) to assemble clusters of biotin-labeled kinesins (KSA). To make K401-streptavidin clusters, 5.7 µL of 6.6 µM streptavidin was mixed with 5 µL of 6.4 µM K401 and 0.5 µL of 5 mM DTT in M2B. This mixture was incubated on ice for 30 minutes.

*DNA nanostars*

We assembled DNA nanostars from individual oligomers using a two-step annealing process. We obtain individual oligomers (Integrated DNA Technologies) with the following sequences:

1. CGCGGATAAATGAAACACCGAACCGCTGTGCAGTAAAGCCCGACGCGCG
2. CGG GCTTTACTGCACAGCGGAACTCGAGAGGTGTCTTGCCGCACGCGCG
3. GCGGCAAGACACCTCTCGAGAACTGGTGAGGCGCAGAGGAACACGCGCG
4. GTTCCTCTGCGCCTCACCAGAACGGTGTTTCATTTATCCGCGACGCGCG

Upon receiving an oligomer shipment, we diluted the oligomers to 1mM in water. We mixed each oligomer, diluting the concentration of each to 250 µM. We then further diluted the oligomer mix with M2B buffer (PIPES) to 180 µM. We added YOYO-1 dye at a volume ratio of 1/100. We prepared all samples in batches of 100 µL.

Once mixed, we annealed the DNA nanostars in a thermocycler. The mixture was first incubated at 95°C for 5 minutes to ensure no oligomers were annealed. We then ramped the temperature down at a rate of -0.5°C/minute until the temperature reached 15°C. We stored the nanostars at 4°C.

We prepared all experimental samples in a flow chamber composed of acrylamide-treated glass slides and coverslips. The acrylamide surface treatment helps prevent protein aggregation. We cleaned each glass slide using Hellmanex and water via sonication for 10 minutes. After cleaning, we ensured all the glass is completely dry. We prepared a Silane-coupling solution composed of 98.5% ethanol, 1% acetic acid, 0.5% 3-(Trimethoxysilyl)propyl methacrylate. We immersed the glass in the silane solution for 10 minutes, then rinsed the glass with water. We then prepared a 2% acrylamide solution in water, and degased for 30 minutes. After degassing, we added 0.7mg/mL ammonium persulfate and 0.03% Tetramethylethylenediamine. We poured the acrylamide solution over the glass slides and coverslips and allowed them to incubate for at least 24 hours. We rinsed the glass slides and coverslips with water and air-dried them before using them for an experiment.

**Active fluid assembly**

We prepared active fluid samples using the same protocols outlined in [39] with the following modifications. We increased the concentration of $MgCl_2$ to 12mM to ensure the DNA nanostars phase-separate at a temperature suitable for microtubule stability. We also remove the M2B buffer from the active fluid "pre-solution" to make room for the DNA nanostars. We prepared large (~500 µL) batches of the active fluid recipe without microtubules, kinesin motor clusters, and DNA nanostars to improve the robustness and repeatability of the experiments.

On the day of all experiments, we mix microtubules (1.33 mg/mL), kinesin motor clusters (120 nM, unless stated otherwise), and DNA nanostars (10 µM, unless stated otherwise) with the other "pre-made" components of the active fluid recipe. We immediately pipette samples to a glass slide chamber to mitigate time for the DNA to phase separate into droplets.

### B. Microscopy and image analysis

Experiments were performed on an inverted motorized Nikon Ti2 microscope, equipped with a motorized stage, a SOLA light engine (Lumencor), and a sCMOS Hamamatsu Orca Flash V4 camera. The illumination and the data acquisition were controlled by µManager (µManager, Version 2.0.0-gamma [53]). All the measurements were performed at room temperature.

*Droplet segmentation and size measurements*
We performed droplet detection with a custom Python script. **Fig. S4** shows the basic steps. We first segmented the maximum gradient projection using a local threshold filter with a window size of 15 pixels. We then skeletonized the foreground of the local threshold detection to get edges of droplets that are 1 pixel in width.  We then filed all closed circles as part of the detection foreground.  We removed any noise by removing objects smaller than 12 pixels from the segmentation mask result. We then performed a watershed algorithm on the segmented droplets before making measurements. Finally, we used the Scikit-Image function regionprops to obtain the total number of pixels (A) in each unique droplet.  We calculated the radius of each droplet as $r = \sqrt{A/\pi}$.

*Tracking droplets*
We carried out droplet tracking using the same experimental setup as the coarsening experiments. We prepared the active fluid with DNA nanostars of varying concentrations to ensure a variety of droplet sizes at the same time.  We quenched the system to a temperature T = 22°C and we allowed droplets to coarsen for approximately four hours.  Once the system was at a low enough number density, we began imaging the motion of the droplets and active fluid. We imaged this experiment with a 4x magnification Nikon objective and a Hamamatsu Orca camera.  We imaged the system once every 15 seconds.  Only one slice in the z-axis was required to image the most trackable droplets. To segment these droplets, we used the Phansalkar local thresholding algorithm with a contrast parameter set at 15 (available in ImageJ).  We then used the FIJI plugin TrackMate[54] to analyze droplet trajectories.  We used the simple LAP tracker with a maximum step size of 25 µm and no gap closing.  We exported the trajectory data to be further analyzed in Python.

We used a custom-written Python routine to further analyze the trajectories.  First, trajectories were filtered by length.  We set the minimum trajectory length to 30,000 seconds, or 2000 times the imaging interval rate.  We calculate the time-averaged mean squared displacement of the droplets up to half of the trajectory length for statistical quality.  We then used the curve_fit function from SciPy's optimize module to fit each curve to the enhanced diffusion model:

$$< \Delta x^2 > = 4D_{eff}\Delta t[1 - e^{-\Delta t/\tau}].$$

We also calculated the standard deviation on the MSD for each lag time and input these values as y-errors on the points used in the fit.

*Particle Image Velocimetry*
We used PIVlab v2.60 to perform analysis on active fluid flows. We used a four-pass FFT window deformation algorithm with window sizes 256, 128, 64, and 32 pixels, with 50% overlap between windows. We used the Gauss 2x3 point sub-pixel estimator function and the standard correlation robustness. After measuring the flow data, we perform a standard deviation threshold (n=6) to remove vectors that are too large to be physically possible in the experiment. We then export data for further analysis in Python. The mean speed is an average over space and time of the velocity field.

To calculate the characteristic correlation length, we used a custom Python script on the output of PIVlab data. We used the correlate2d function from the SciPi module to measure correlations in the x and y components of flow velocity. The total correlation was computed as Cx + Cy. We then computed the average correlation as a function of radial distance, C(r). We then fit this function to an exponential form: $C(r) = e^{-r/\lambda}$, where λ is the characteristic correlation length.

*Method to identify the bounds to fit the power law for N(t) and r(t) (Experiments)*
Fit ranges for experimental data were bounded by experimental constraints at both short and long times. At short timescales, we began fitting when the average radius of droplets is at least $0.7\mu m$. This size corresponds to when droplets are large enough to accurately measure (more than 13 pixels) given the 20x microscope objective used in all coarsening experiments. At late times, we end the fit when we started detecting finite size effects on the coarsening dynamics. This corresponds to the time when the radius of the largest droplet in the FOV stars plateaus.

*Method to identify the bounds to fit the power law for N(t) and r(t) (Simulations)*
Fits to power law scalings in simulations were bounded to exclude non-scaling regions in the curves for $\bar{R}(t)$ and $N(t)$. The lower bound, where fitting began, was calculated to start when $\bar{R} = 1$. This coincided with when the scaling behavior started, indicated by a straight line with log-log scaling. The fit region ended when the maximum radius reached 95% of the maximum radius possible in the system, given by the total volume fraction. This is roughly when the droplets begin showing finite number effects, near the end of the simulation (**Fig. S17**). If this maximum value is never reached in the simulation, the fit is carried out until the last time point $t = 10^4$.

### C. Details about the computer simulations

Droplet simulations were all carried out using a custom-written MATLAB (R2020b) routine. Droplets were initialized in a box of size 200x200x40 for both active and passive fluid simulations. The box size was chosen to match the dimensions of the experimental flow chamber. Droplet locations and radii were generated such that no droplets overlap at the start of the simulation. The radii of droplets at the start of the simulation were drawn from a normal distribution of average size 0.8 and a standard deviation 0.08.

In the case of passive fluid simulations, we generated trajectories using Matlab's wfbm() with a Hurst parameter H = 0.5. Trajectories generated were then only accepted for the simulation if they result in a diffusive exponent of α = 1 +/- 0.05. At each time point during the simulation, we update the position of each droplet according to the pre-generated trajectory assigned to that droplet. Anomalously diffusive droplet trajectories were generated for a Hurst exponent H= 1 and H=0.25 . We measured the MSD to confirm that the diffusive exponent (α) for H = 1 and H = 0.25 give $\alpha = 2$ and $\alpha = 0.5$, respectively.

*Hydrodynamic simulations*

We simulate an incompressible active nematic flow V using a previously published minimal 3D nematohydrodynamic model [55]. The box dimension was $L_x=L_y=200$ and $L_z=40$ with periodic boundary conditions along the X and Y axes. A typical example of the flow field in the middle of the channel is shown in **Fig. S8A**. Given the calculated V(**x**,**y**,**z**,t), we seeded the simulation box with 6,000 droplets with a size distribution that matches the experimental one (**Fig. S17A**). Given the low volume fraction of the droplets (**Fig. S3C**), we assumed that the flows are not impacted by the presence of droplets and that the droplets are simply advected by the active flows. We registered droplets' collisions and assumed that overlapping droplets instantaneously coalesce and are replaced by a larger droplet that conserves the initial volumes and is located at their center of mass.

*Active particles simulations*

To generate AOUP trajectories, we solve the set of overdamped Langevin equations:

$$\dot{r}_i = v_i,$$

$$\dot{v}_i = -\frac{v_i}{\tau} + (2D/\tau)^{1/2}\xi_i$$

where r_i is the position of the i-th droplet, v_i is the self-propulsion velocity of the i-th droplet, τ is a persistence time that sets the timescale for the self-propulsion velocity to change, D is a diffusion constant that sets the particle speed, and ξ_i is a Gaussian white noise with mean zero and correlations:

$$< \xi_{i,\mu}(t), \xi_{j,\nu}(t') > = \delta_{ij}\delta_{\mu\nu}\delta(t - t'),$$

where µ and ν index spatial dimensions and i and j index particles. We solve Eqs. 1a,1b numerically using standard Euler-Maruyama integration, using a timestep of $\Delta t = 10^{-3}$ and NumPy's "random.normal" function to generate the requisite random numbers. For all of our simulations, we set $D = 20$ and $\tau = 5$, and to obtain our measurements of mean droplet size versus tim,e we averaged over 6000 trajectories with different random seeds. Our code is available on GitHub at [insert URL here].

ABP dynamics are given by:

$$\dot{r}_i = v_i,$$
$$\dot{v}_i = (2D_r)^{1/2} \xi_i \times v_i,$$

where $r_i$ is the position of the $i$-th droplet, $v_i$ is the self-propulsion velocity of the $i$-th droplet, Dr is a rotational diffusion constant that sets the reorientation time of the self-propulsion velocity, and ξi is a Gaussian white noise with the same statistics as for the AOUPs. The magnitude v0 of the self-propulsion velocity is conserved in the continuous-time limit, but for simulations with a finite timestep, small errors will accumulate and cause the magnitude of $v_i$ to change. We therefore re-normalize $v_i$ to have a magnitude $v_0$ after every timestep. We set $D_r$ = 0.2 and $v_0$ = 3.

Finally, for RTPs, the position $r_i$ evolves according to the self-propulsion velocity $v_i$ as before:

$$\dot{r}_i = v_i,$$

The self-propulsion velocity has a constant magnitude $v_{RTP}$, but switches orientation on a timescale $\tau_r$. Specifically, at each timestep, with probability $1 - e^{-\Delta t/\tau_r}$, the self-propulsion velocity assumes a new randomly chosen orientation. We set $v_{RTP}$ = 3 and $\tau_r$ = 5.

In the case of active fluid simulations, droplets do not use pre-generated trajectories. Rather, we have pre-generated vector fields resulting from active fluid simulations, using the same method as [55] with H = 40 and W = 200. At each time point, the corresponding flow data is loaded, and droplets are displaced by an amount that corresponds to the nearest vector to the center of the droplet. Droplets do not interact or change the flows of the active fluid regardless of their size. Unless otherwise stated, there is no Stokes-Einstein relationship between the effective diffusivity of droplets and the droplet size, as is present in the passive fluid simulation.

Periodic boundary conditions were kept the same for both types of simulations. Unless otherwise specified, all simulations were run over a period of $10^4$ steps. Each simulation condition was repeated a total of at least 12 times. We used the Parallel Computing Toolbox to run up to six simulations in parallel.

To measure the droplet collision kernels, the initial distribution of droplets was chosen to be uniform between radii 0.8 and 8, with an equal number of droplets in each logarithmically spaced bin. Furthermore, we didn't allow droplets to merge but instead collected information about which droplets overlap throughout the simulation.

## D. Supplementary figures

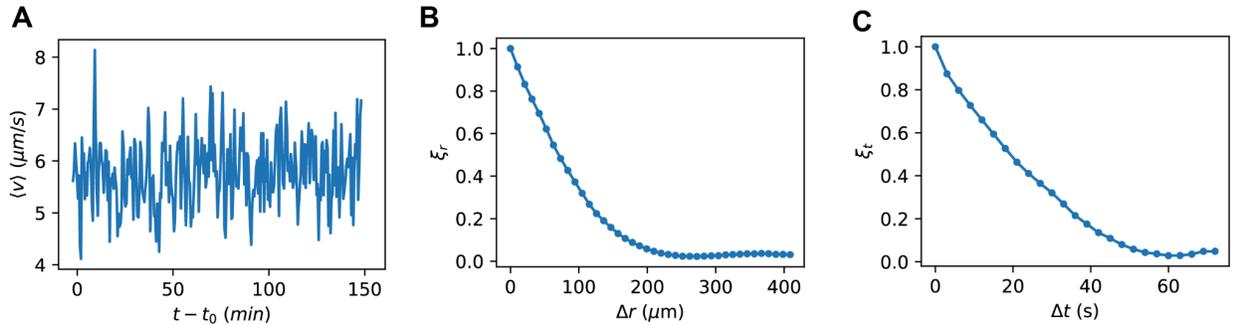

**Fig S1: Characterization of the self-stirring active flows**. A) mean speed versus time. B) Spatial velocity autocorrelation function. C) Temporal velocity correlation function.

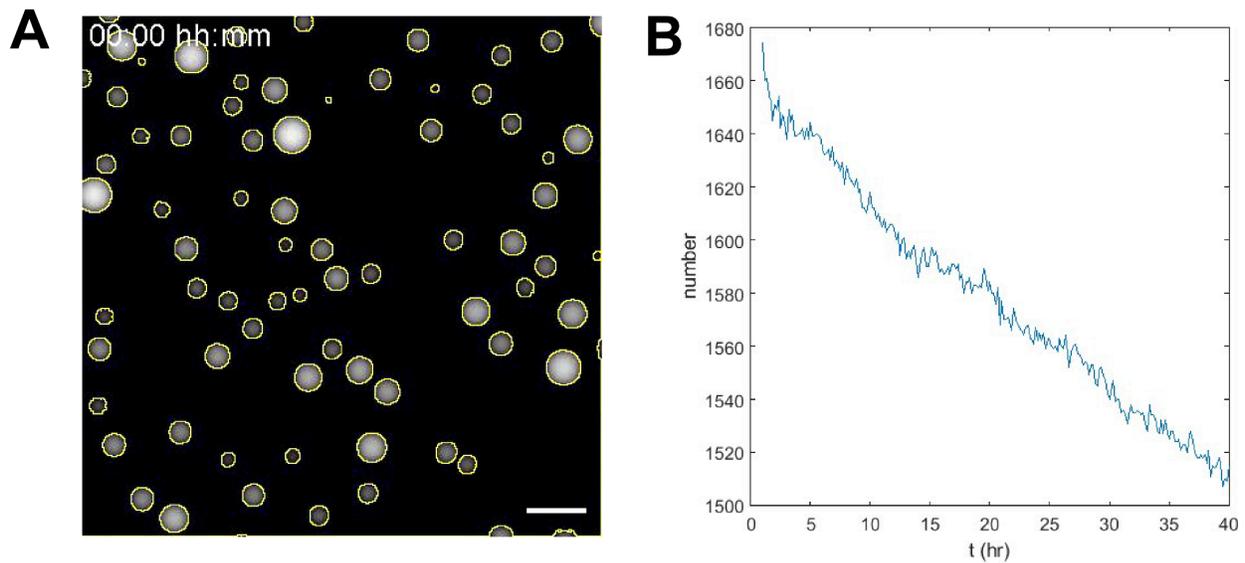

**Fig. S2: Ostwald ripening in immobile droplets.** A) Snapshot of immobile droplets immersed in a passive fluid. B) The total number of droplets decreases over time significantly slower than when the droplets diffuse or are advected by an active fluid (**Fig. 1L**)

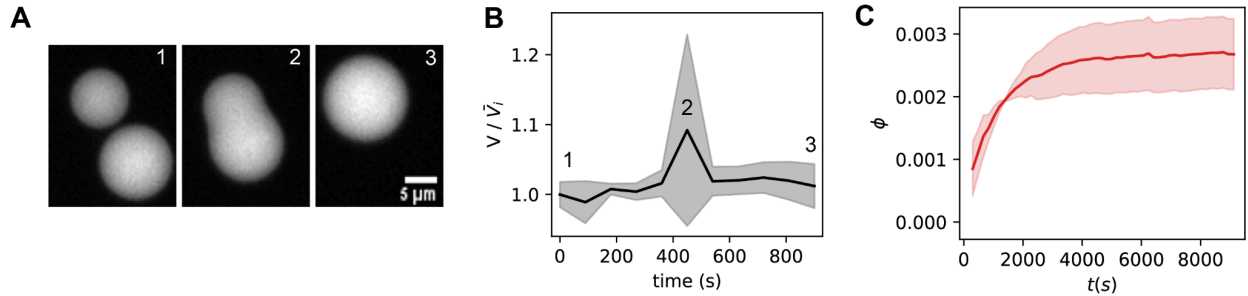

**Fig. S3: Droplet coalescence drives coarsening**. A) A time series of two droplets coalescing (400 s between each frame). B) Rescaled volume $V/\tilde{V}$ of colliding droplets as a function of time. When the droplets have a spherical shape, $V/\tilde{V}=1$. Collision occurs at t=400sec. Upon collision, $V/\tilde{V}$ increases and then relaxes to 1 as the droplet relaxes towards a spherical shape (N = 7 pairs of droplets). C) The volume fraction reached a stable value within 2000 sec. Volume fraction is inferred from the total surface of the segmented DNA-rich droplets.

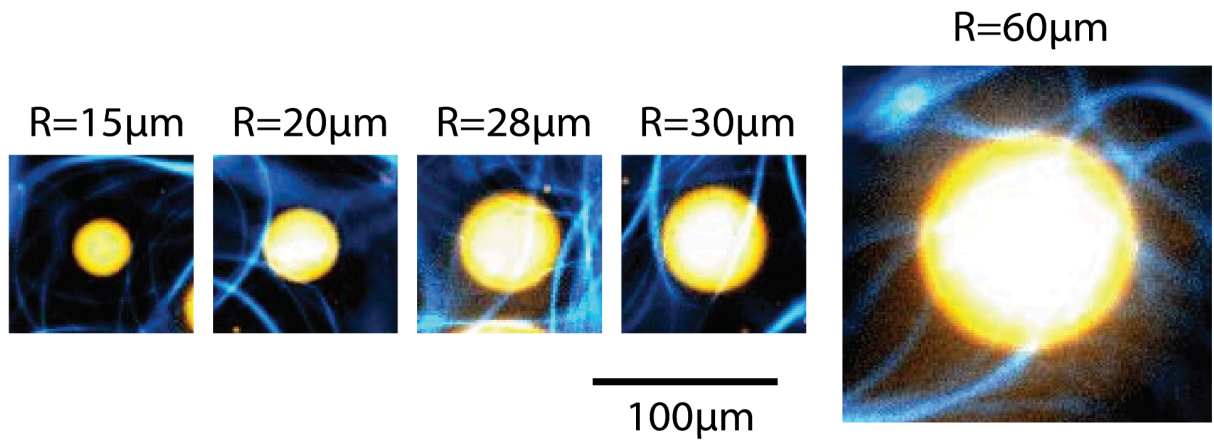

**Fig. S4: The DNA-rich droplets are not constrained by the microtubule network**. Representative images of DNA-rich droplets (yellow) suspended in a self-stirring microtubule network (cyan).

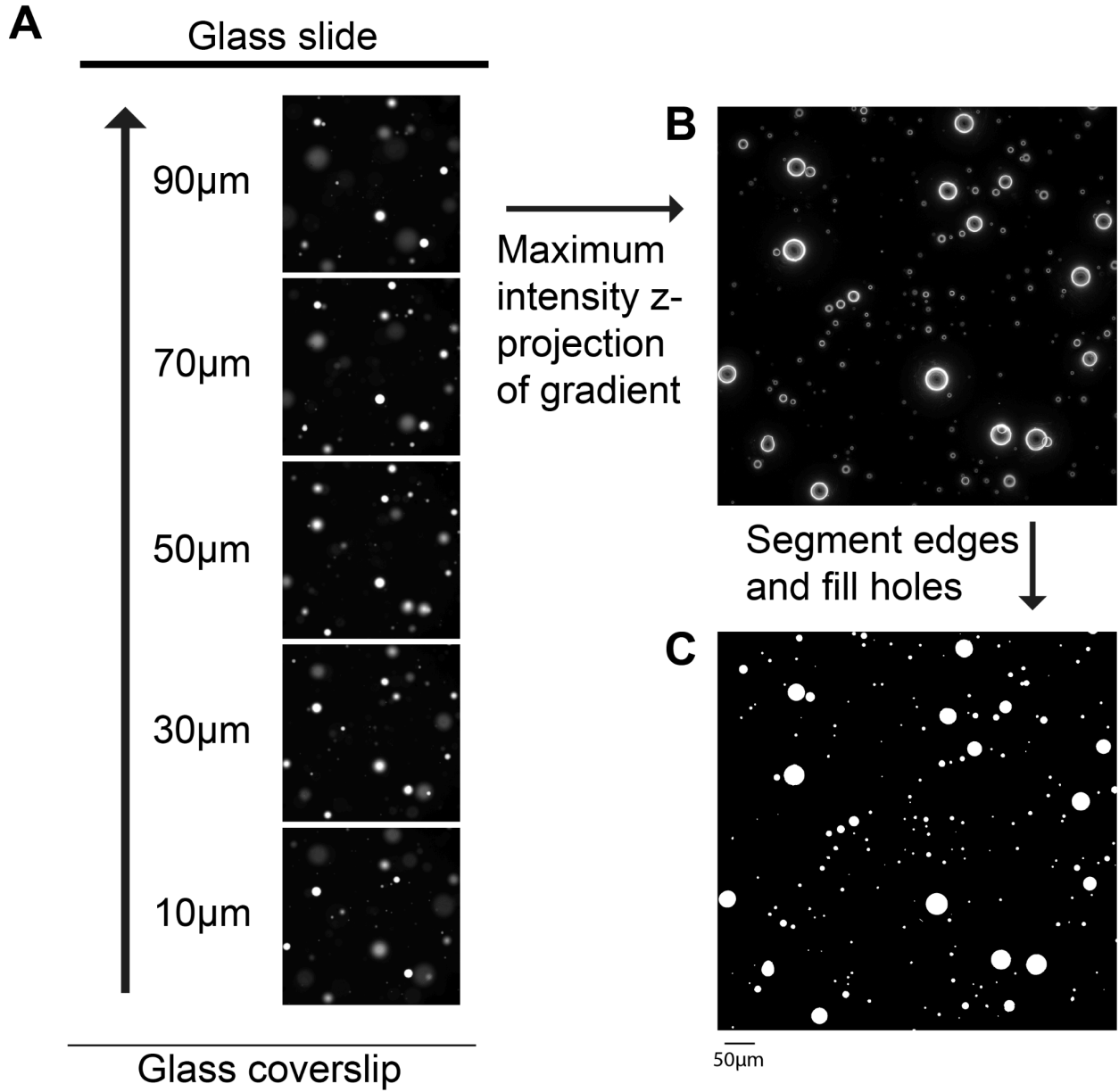

**Fig. S5: Image analysis pipeline to accurately measure the droplet size distribution.** A) For each X-Y position, a stack of images spaced by 20 µm is taken. B) We performed a maximum intensity Z-projection of the gradient of the intensity. C) We segmented the image in B and filled the hole to detect the droplets.

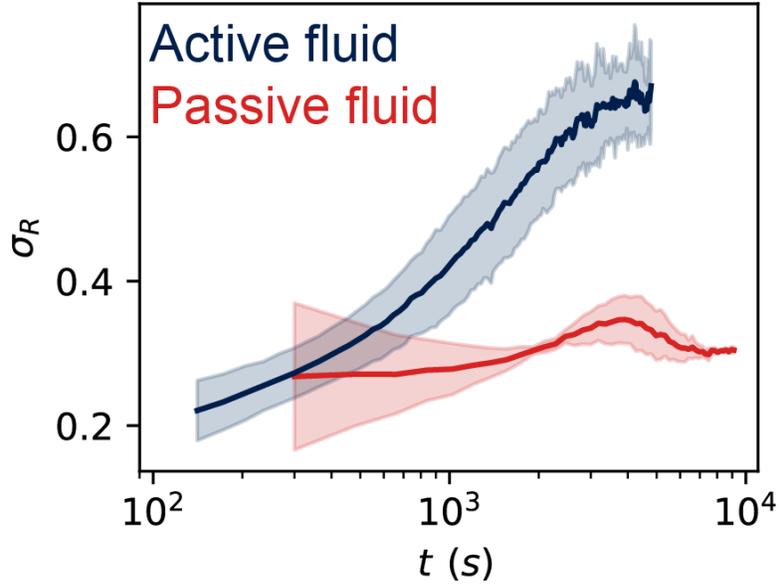

**Fig S6: Second moment of the log-normal size distribution $\sigma_R$ for droplets in an active fluid (blue) and in a passive fluid (red)**

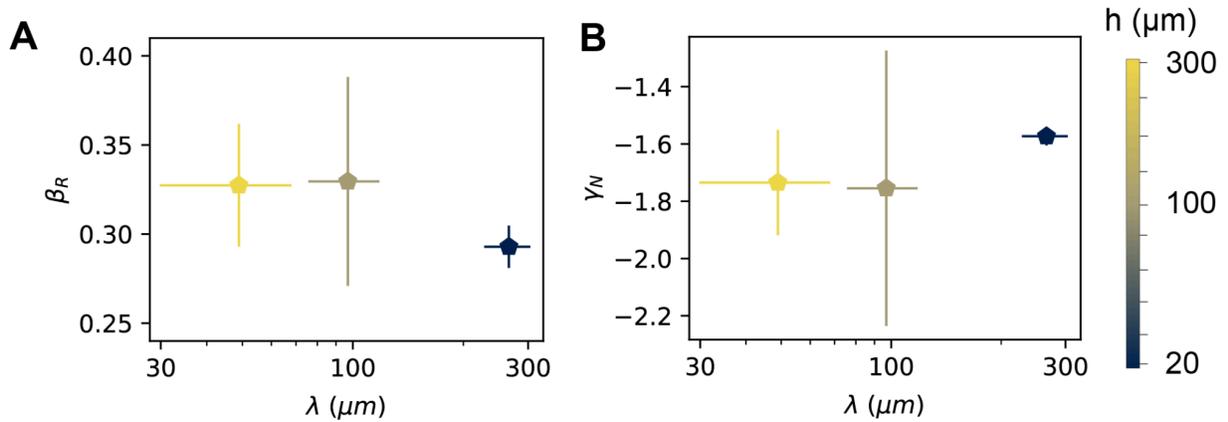

**Fig S7: Coarsening exponents are independent of the active length scale λ** which is controlled by changing confinement along the Z axis [57]. A) $\beta_R$ vs. λ, $\gamma_{Nu}$ vs. λ. The high of the flow chamber is color-coded.

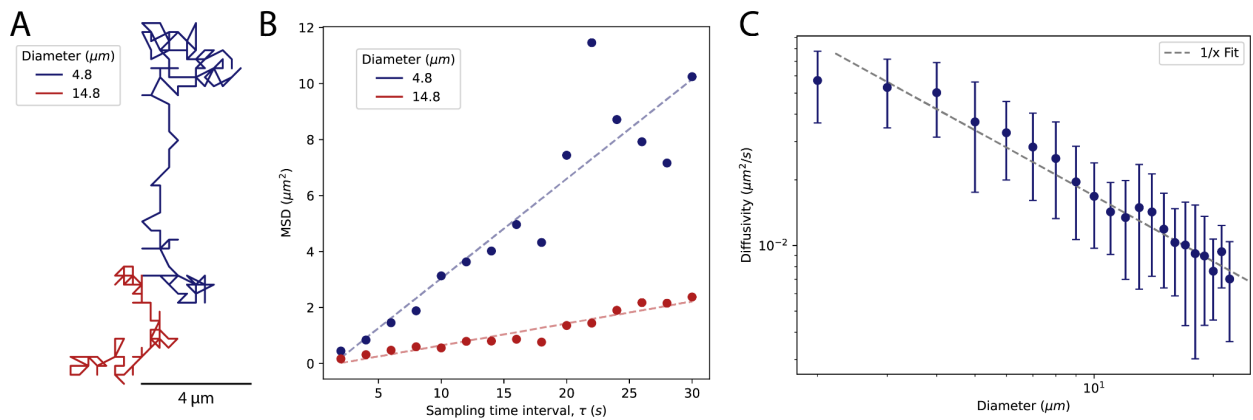

**Fig. S8: Passive droplets respect the Stokes-Einstein-Sutherland relationship.** A) Trajectories of two passive droplets of different diameters. B) Mean Squared Displacement of the droplets shown in A. C) Log-log plot of the diffusivity of the droplets as a function of the droplet's diameter. Experimental points and error bars shown in blue correspond to means and standard deviations; the theoretical prediction for the Stokes-Einstein-Sutherland relation is shown as a dashed line.

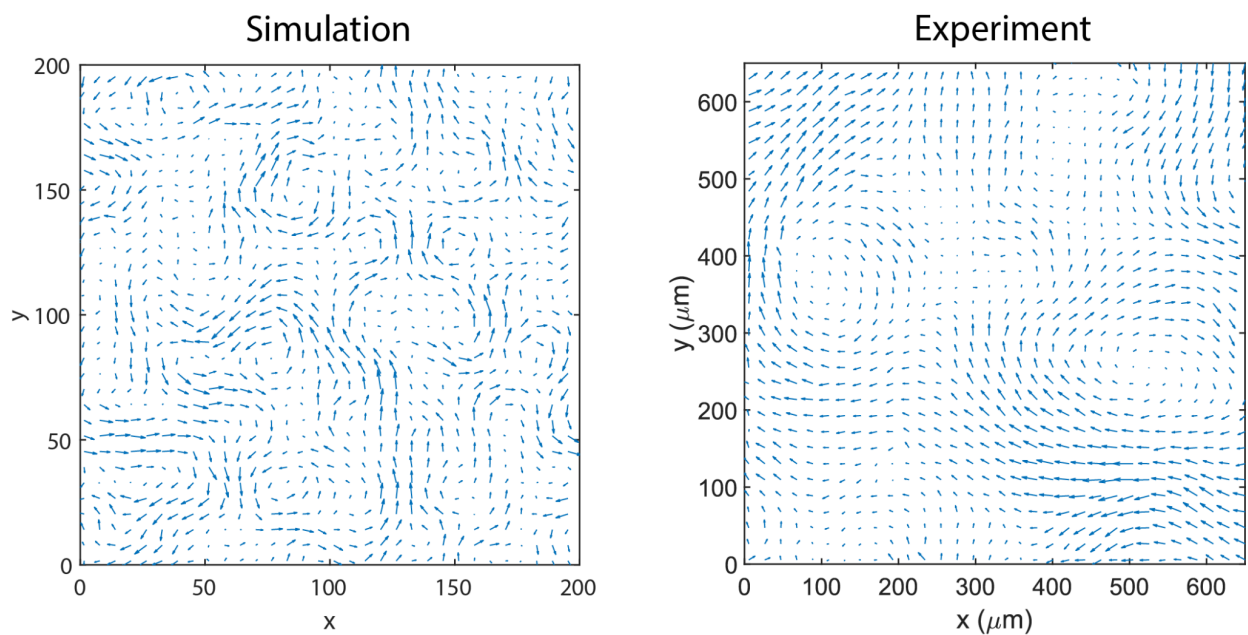

**Fig. S9: In-plane flow fields in simulation (left) and experiments (right).**

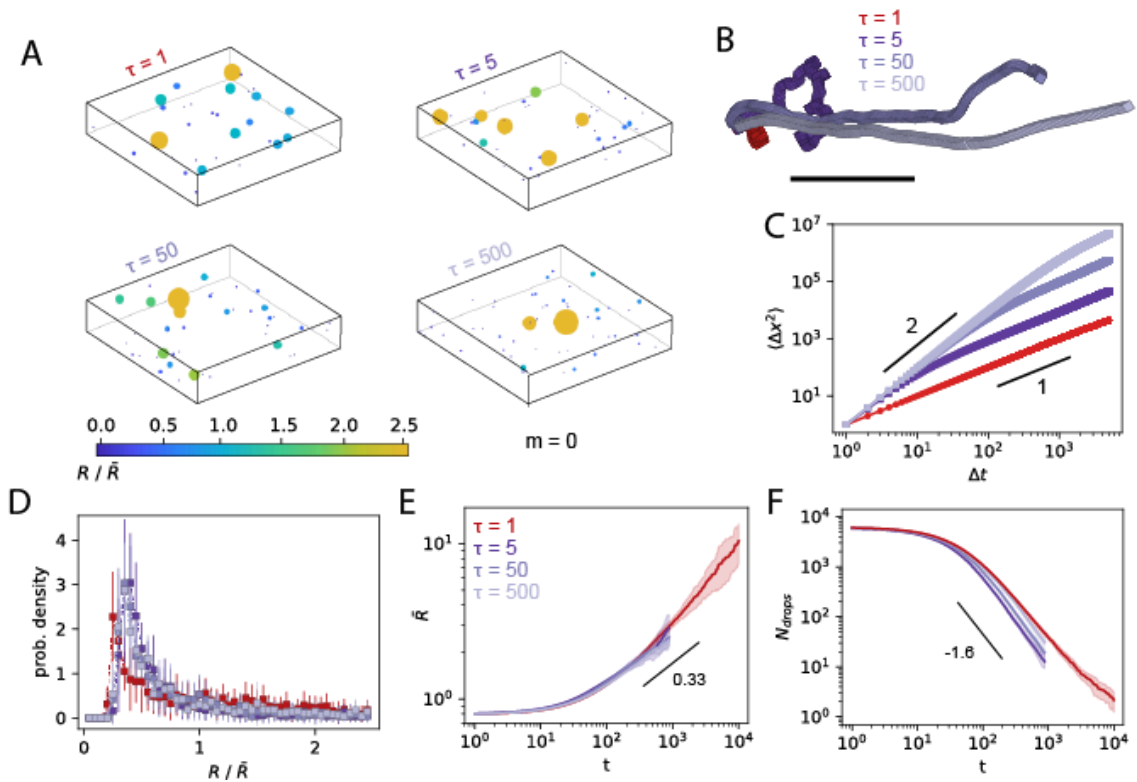

**Fig. S10: Coarsening dynamics for self-propelled AOUPs with various persistence time τ.**
A) Snapshots of the simulations for coarsening AOUPs with persistence time varying from τ=1 to τ=500. Simulations started with the same number of droplets with the same size distribution. Each simulation box shown here has the same total number of droplets. Droplets are color-coded according to their relative size R/R̄, where R̄ is the mean droplet radius. B) Characteristic trajectories of the AOUPs with different persistence times (scale bar:100). C) Mean Square Displacement of the AOUPs. D) Probability distribution function of droplet's radius. E) Mean droplet radius as a function of time. F) Total number of droplets as a function of time.

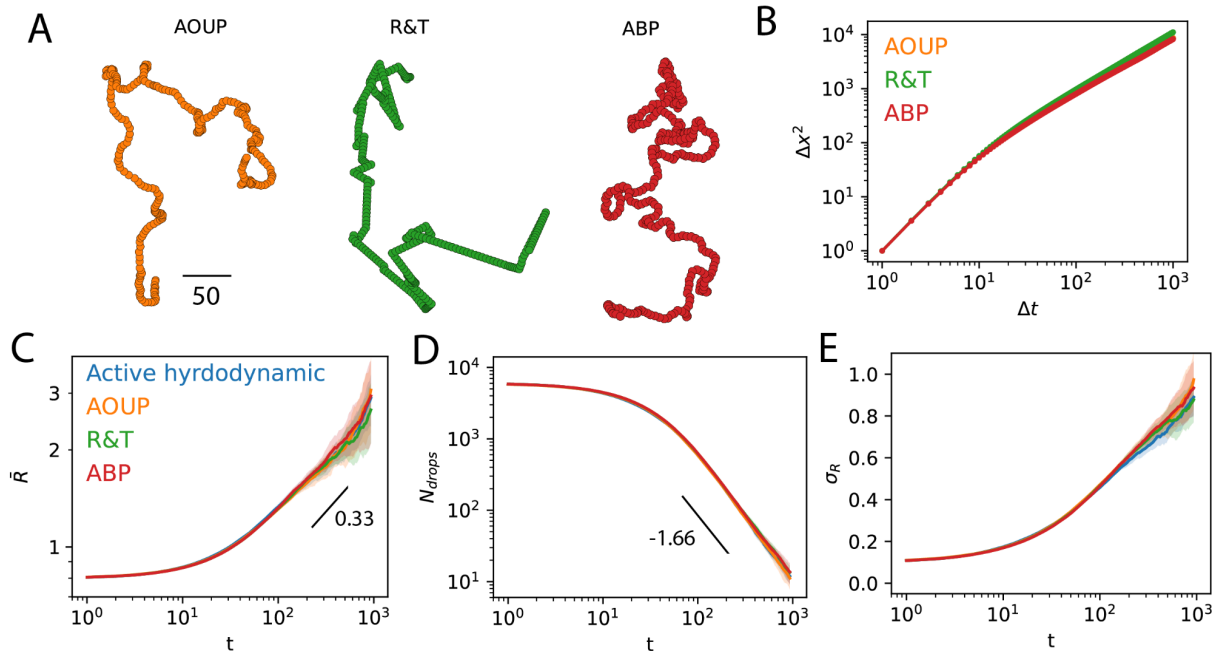

**Fig. S11: Comparing the coarsening of AOUPs, Run-and-Tumble (R&T) Particles, and Active Brownian Particles (ABPs).** A) Characteristic trajectories. B) Mean Square Displacement of the self-propelled particles. C) Mean droplet radius as a function of time. All the lines are superimposed on top of each other. D) Total number of droplets as a function of time. E) Second moment of the log-normal size distribution as a function of time.

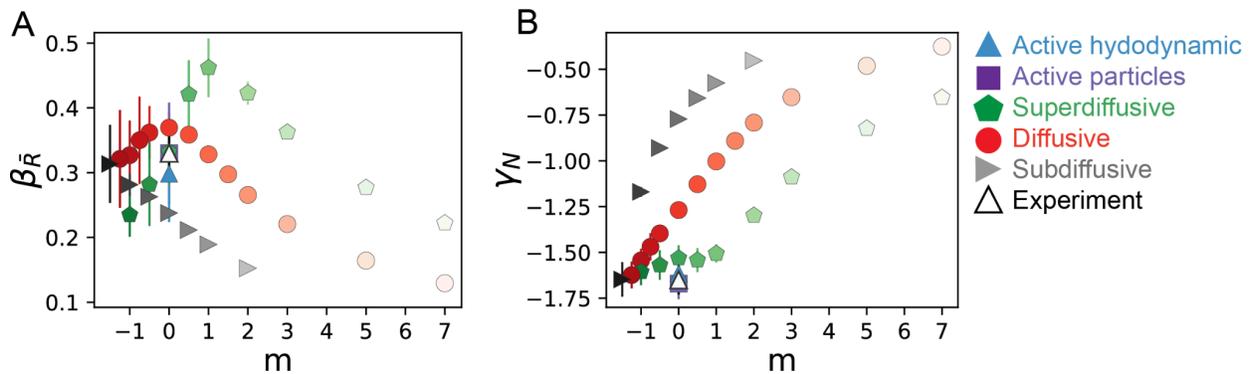

**Fig. S12: Coarsening exponents for anomalous diffusion.** Coarsening exponents for A) the mean radius and B) the total number of droplets as a function of the SES scaling exponent m ($D \sim R^{-m}$, where D is the diffusivity and R is the droplet's radius).

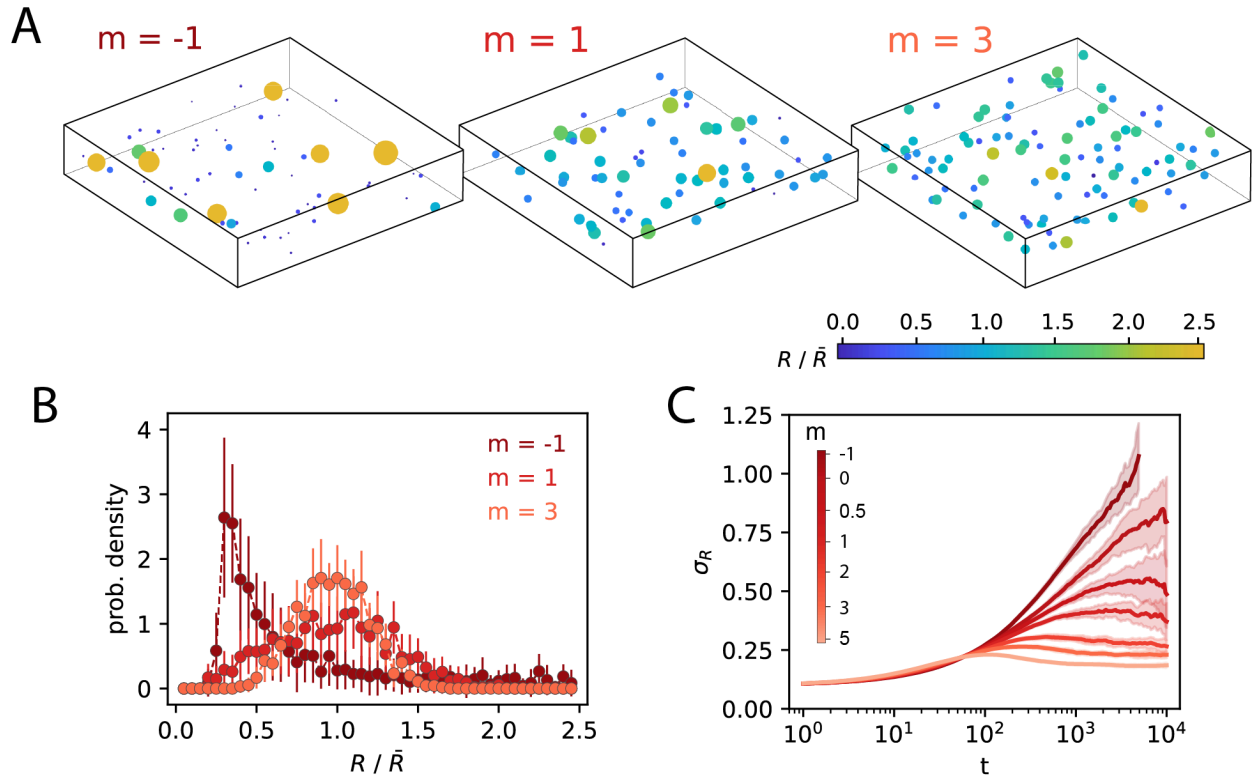

**Fig. S13: Impact of the Stokes-Einstein-Sutherland scaling on the coarsening dynamics.**
A) Snapshots of the simulations for coarsening of diffusive particles with a modular SES scaling exponent m ($D \sim R^{-m}$, where D is the diffusivity and R is the droplet's radius). Simulations started with the same number of droplets with the same size distribution. Each simulation box shown here has the same total number of droplets. Droplets are color-coded according to their relative size $R/\bar{R}$, where $\bar{R}$ is the mean droplet radius. B) Probability distribution function of droplet's radius for simulations containing the same number of droplets. C) Second moment of the size distribution as a function of time for various values of m.

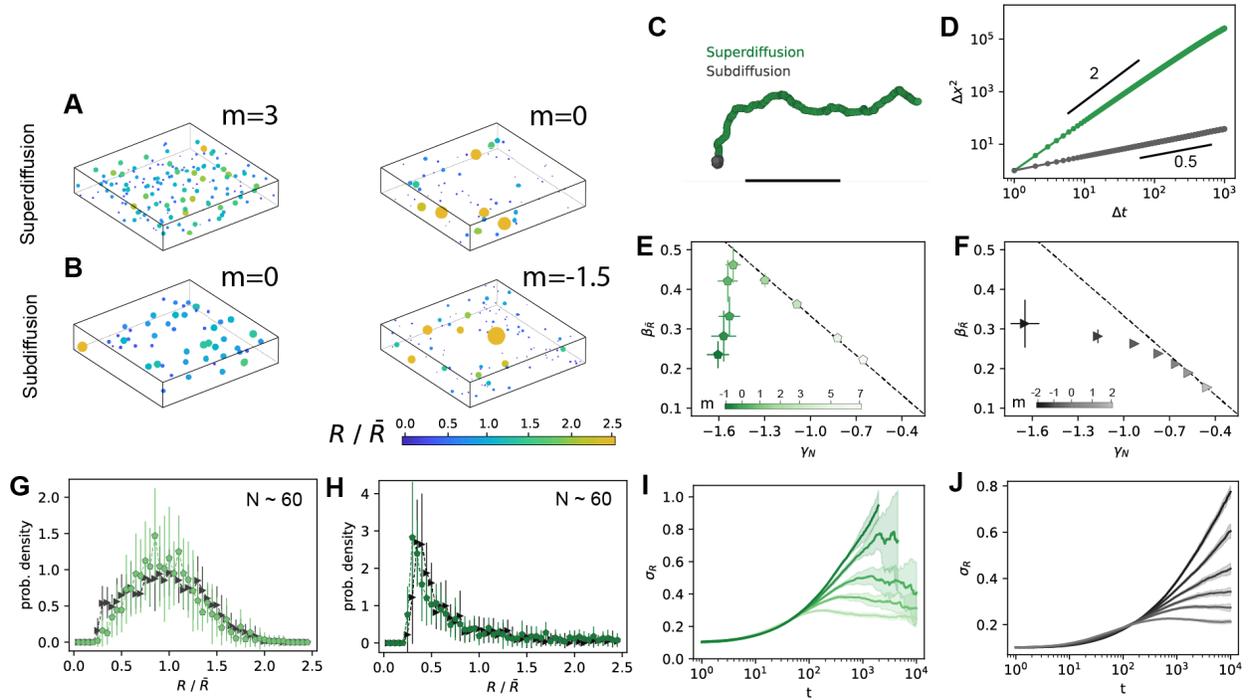

**Fig S14: Impact of anomalous diffusion exponent on the coarsening dynamics.** A-B) Snapshots of the simulations for coarsening of A) superdiffusive ($\alpha$=1.67) and B) subdiffusive particles ($\alpha$=0.5) with a modular Stokes-Einstein exponent m. m was chosen so that the collision kernel exponent is equal to 0 (let panels) or 2 (right panels). Droplets are color-coded according to their relative size $R/\tilde{R}$, where $\tilde{R}$ is the mean droplet radius. C) Example of trajectories for anomalously diffusing droplets. D) Mean Square Displacement vs. time for the superdiffusive and the subdiffusive droplets. E-F) $\beta_R$ and $\gamma_N$ coarsening exponents for E) superdiffusive and F) subdiffusive droplets with a modular Stokes-Einstein-Sutherland's exponent m ($D\sim R^{-m}$). G-H) relative size distribution function for G) k=0 and H) k=2. I-J) Second moment of the size distribution as a function of time for I) superdiffusive and J) subdiffusive droplets.

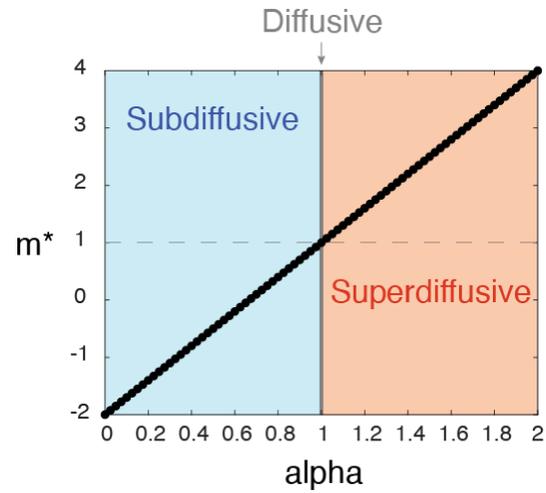

**Fig. S15: Linear relationship between m* and α.** m* is the value of SES exponent m for which k=0. For m>m*, size distributions are self-similar and $\beta_R = -\gamma_N/3$. For m<m*, size distributions are not self-similar, and the relationship between $\beta_R$ and $\gamma_N$ is non-linear.

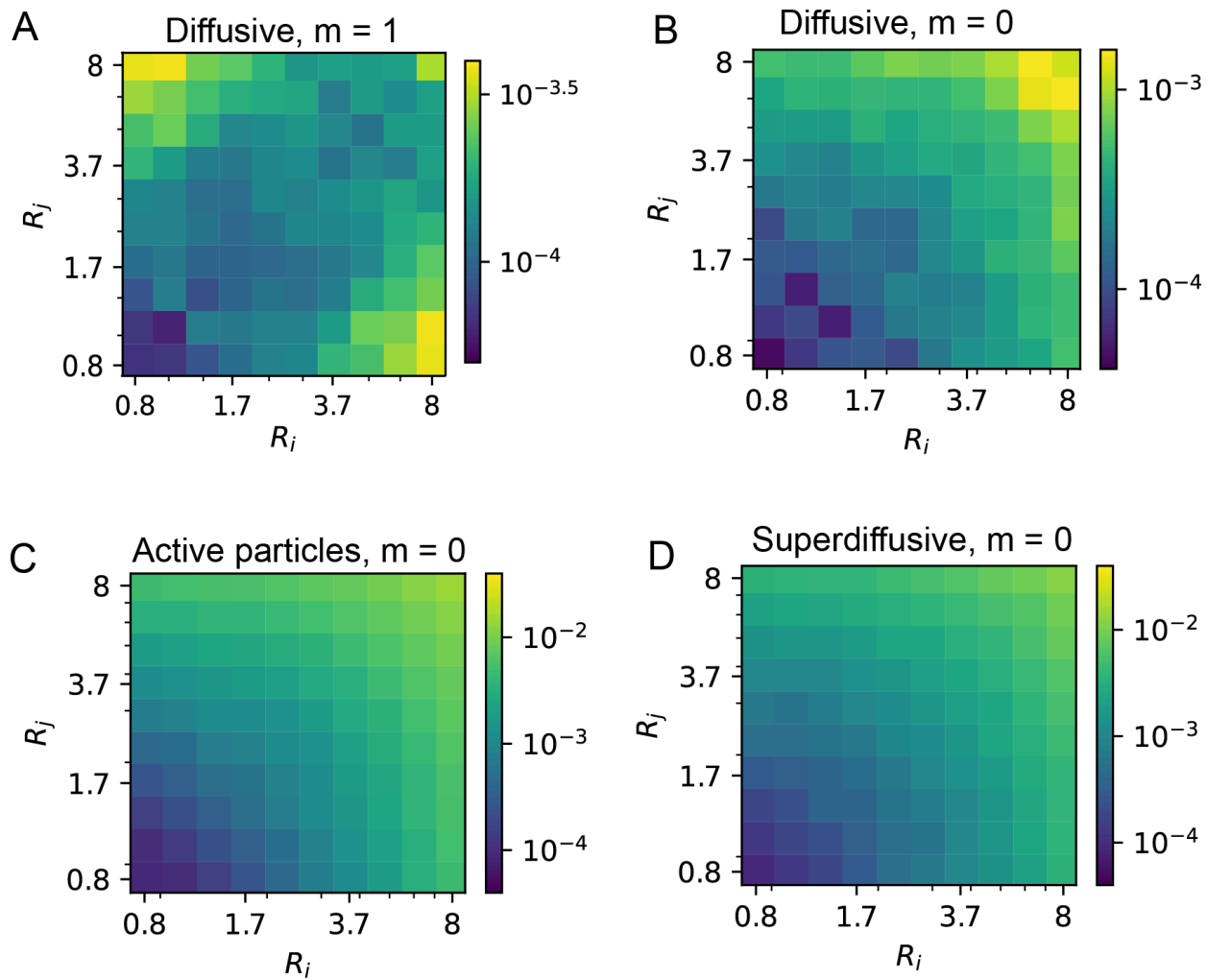

**Fig. S16: Collision kernels** for A) Brownian particles (α=1, m=1), B) diffusive particles where diffusivity is size-independent (α=1, m=0), C) AOUPs with size-independent diffusivity(m=0), and D) Superdiffusive particles with size-independent diffusivity(α=1.67, m=0).

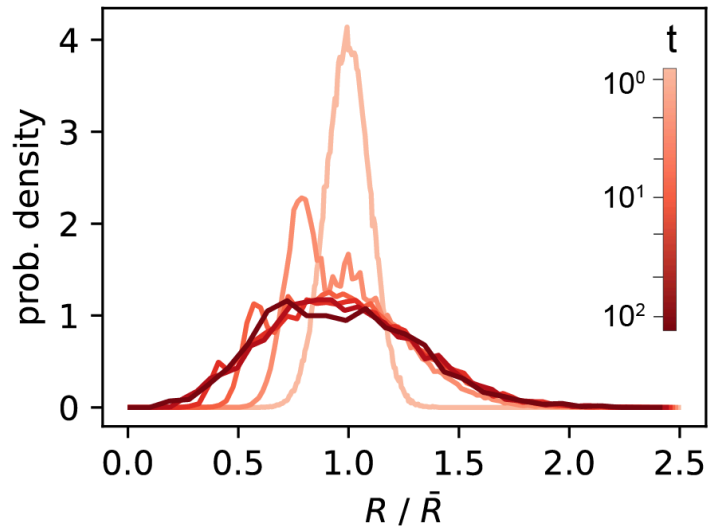

**Fig. S17: Self-similarity in the probabilistic calculation for a Brownian collision kernel.** Temporal evolution of the probability distribution of the droplets' relative radius $R/\bar{R}$, where $\bar{R}$ is the mean droplet radius, for a collision kernel exponent $k=0$ (Smolukowski kernels for brownian droplets).

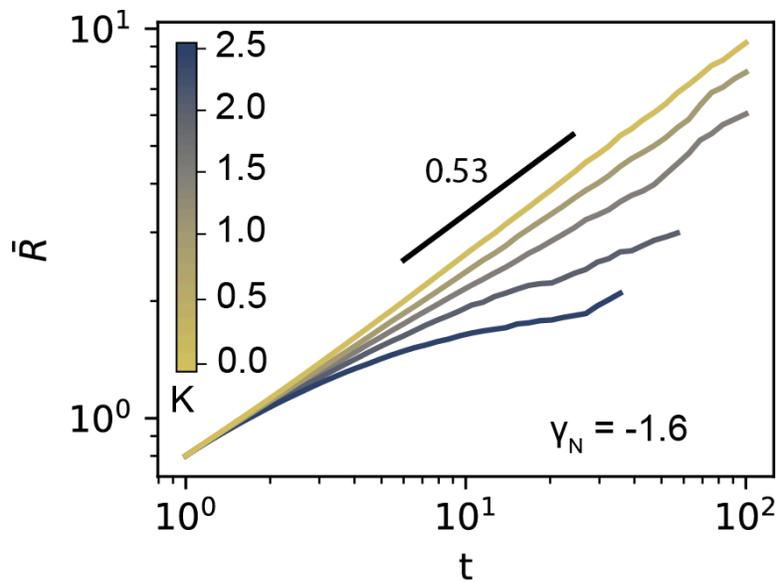

**Fig. S18: Impact of the collision kernel exponent k on the coarsening dynamics**. Temporal evolution of the average droplet's radius $\bar{R}$ for various values of k and a prescribed $\gamma_N$=-1.6.

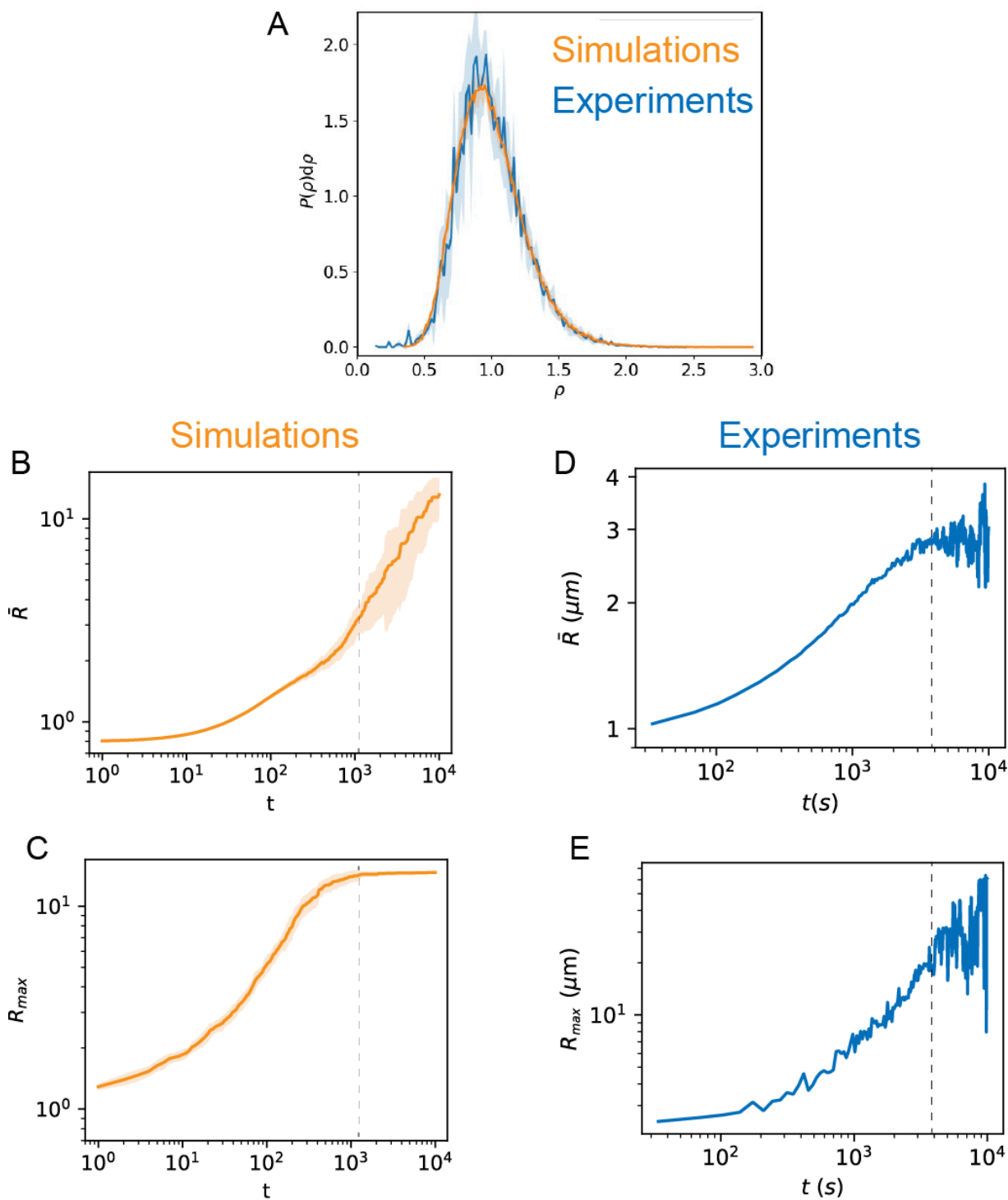

**Fig. S19: Finite number effects limit the fitting range.** A) Probability distribution of the droplets' relative radius R/R̄, where R̄ is the mean droplet radius, in experiments and simulations. B) Mean radius R̄ vs. time and C) radius of the largest droplet vs time for the active hydrodynamic coarsening simulation. D) Mean radius R̄ vs. time and E) radius of the largest

droplet over time for one example of an active fluid coarsening experiment ([ATP]=1400 μM, h=100 μm).

### E. Supplementary movie description

**Movie S1: Coarsening of DNA droplets embedded in an active fluid.**
**Movie S2: Tracking droplet trajectories advected by the active flows.**
**Movie S2: Simulations of droplet coarsening in an active fluid.**
**Movie S3: Simulations of coarsening for diffusive droplets with m=1, m=0, and m=-1.**